\documentclass[preprint,12pt]{elsarticle}

\usepackage{amssymb}

\journal{Nuclear Physics A}

\begin{document}

\begin{frontmatter}



\title{Two-Baryon Potentials and $H$-Dibaryon \\ from 3-flavor Lattice QCD Simulations}



\author[label1]{Takashi Inoue}

\author[label2,label3]{Sinya Aoki}

\author[label4]{Takumi Doi}

\author[label5,label7]{Tetsuo Hatsuda}

\author[label8]{Yoichi Ikeda}

\author[label3]{Noriyoshi Ishii}

\author[label7]{Keiko Murano}

\author[label3]{Hidekatsu Nemura}

\author[label3]{Kenji Sasaki}

\author{(HAL QCD Collaboration)}

\address[label1]{Nihon University, College of Bioresource Sciences, Kanagawa 252-0880, Japan}

\address[label2]{Graduate School of Pure and Applied Sciences, University of Tsukuba, Ibaraki 305-8571, Japan}

\address[label3]{Center for Computational Sciences, University of Tsukuba, Ibaraki 305-8571, Japan}

\address[label4]{Center for Nuclear Study, The University of Tokyo, Tokyo 113-0033, Japan}

\address[label5]{Department of Physics, The University of Tokyo, Tokyo 113-0033, Japan}


\address[label7]{Theoretical Research Division, Nishina Center, RIKEN, Saitama 351-0198, Japan}

\address[label8]{Department of Physics, Tokyo Institute of Technology, Tokyo 152-8550, Japan}


\begin{abstract}

Baryon-baryon potentials are obtained from 3-flavor QCD simulations
with the lattice volume $L \simeq 4$ fm, the lattice spacing $a\simeq 0.12$ fm,
and the pseudo-scalar-meson mass $M_{\rm ps} =469$ -- $1171$ MeV.
The $N\!N$ scattering phase-shifts and the mass of $H$-dibaryon in the flavor $SU(3)$ limit 
are extracted from the resultant potentials by solving the  Schr\"{o}dinger equation.
The $NN$ phase-shifts in the $SU(3)$ limit is shown to have qualitatively similar behavior as the experimental data. 
A bound $H$-dibaryon in the $SU(3)$ limit is found to exist in the flavor-singlet $J^P=0^+$ channel
with the binding energy of about 26 MeV for the lightest quark mass  $M_{\rm ps}=469$ MeV. 
Effect of flavor $SU(3)$ symmetry breaking on the $H$-dibaryon is estimated
by solving the coupled-channel Schr\"{o}dinger equation for $\Lambda\Lambda$-$N\Xi$-$\Sigma\Sigma$ with the physical baryon masses
and the potential matrix obtained in the $SU(3)$ limit:
a resonant $H$-dibaryon is found between $\Lambda\Lambda$ and $N\Xi$ thresholds in this treatment.     
    
\end{abstract}


\end{frontmatter}


\section{Introduction}
\label{sec:intro}

The nucleon-nucleon ($NN$), hyperon-nucleon ($YN$) and hyperon-hyperon ($YY$)
interactions are crucial not only for understanding ordinary nuclei, hypernuclei and
the neutron-star interiors~\cite{Hashimoto:2006aw,SchaffnerBielich:2010am} but also for exploring exotic multi-quark systems
such as the $H$-dibaryon ~\cite{Jaffe:1976yi,Sakai:1999qm,Takahashi:2001nm,Yoon:2007aq}.

There are two methods proposed so far to study  hadron interactions from 
first principle lattice QCD simulations:  
the L\"uscher's finite volume method for scattering phase-shifts~\cite{Luscher:1990ux}
and the HAL QCD method for hadron-hadron potentials~\cite{Ishii:2006ec,Aoki:2009ji}.   
Both methods are  extensively applied to the meson-meson, meson-baryon and 
baryon-baryon systems in recent years and have been shown to be useful to 
derive scattering and bound-state observables. 
In the L\"uscher's method, effect of the finite lattice volume is utilized 
to extract the scattering phase shifts directly~\cite{Fukugita:1994ve,Beane:2002nu,Beane:2010em}.
In the HAL QCD method, the notion of the ``potential", which is insenstive to the finite lattice volume, is introduced
as a kernel of the scattering T-matrix to calculate the observbles~\cite{Nemura:2008sp,Inoue:2010hs,Inoue:2010es,Ikeda:2011qm,Kawanai:2010ru}.

In our previous paper~\cite{Inoue:2010hs},
we reported an exploratory study of the  baryon-baryon ($BB$) potentials by full QCD simulations 
in the flavor $SU(3)$ limit with a lattice volume $L =1.93$ fm and a pseudo-scalar meson mass $M_{\rm ps}= 835, 1014 $ MeV:
We have observed
(i) flavor-spin dependence of the short range part of the $BB$ potentials 
and its connection to the quark Pauli principle, and  
(ii) possible existence of a shallow six-quark bound state in the flavor-single channel
due to attractive core in this channel. 
The latter observation  was further strengthened by our simulations in the flavor $SU(3)$ limit
with a larger volume ($L= 3.87$ fm) and a smaller mass $M_{\rm ps} = 672$ MeV~\cite{Inoue:2010es}
and by an independent study with the (2+1)-flavor simulations
with the pion mass $m_{\pi}=389$ MeV analyzed by the  L\"uscher's method~\cite{Beane:2010hg}. 

One of the purposes of the present paper is to present a full account of the 
$BB$ potentials and $H$-dibaryon on the basis of the lattice data at $L= 3.87$ fm
with the pseudo-scalar meson mass as small as $M_{\rm ps}= 469$ MeV. 
We note that an  extended HAL QCD method of extracting the $BB$ potential from the {\it time-dependent}
Nambu-Bethe-Salpeter wave function~\cite{Ishii:2011} is employed in the present paper
as well as in our recent reports~\cite{Inoue:2010es,Inoue:2011tk}.
Another purpose of this paper is to pursue phenomenological estimate of
the flavor $SU(3)$ breaking on the $H$-dibaryon by taking into account the mass splitting of octet baryons.
In contrast to the recent studies along this line using chiral effective
field theory~\cite{Beane:2011xf,Shanahan:2011su,Haidenbauer:2011ah},
we fully utilize the $BB$ potentials obtained in our simulations and
solve the coupled channel $\Lambda\Lambda$-$N\Xi$-$\Sigma\Sigma$ system
to obtain the phase-shifts and complex poles. 
We find the $H$-dibaryon in the $\Lambda\Lambda$ continuum, a few MeV below the $\Xi\!N$ threshold.
Similar result has been reported before in a coupled channel analysis
of the phenomenological quark model with flavor $SU(3)$ breaking~\cite{Oka:1983ku}.

This paper is organized as follows.
In the next section, we briefly review the extended HAL QCD 
method to study multi-hadron systems in lattice QCD~\cite{Inoue:2010es,Ishii:2011,Inoue:2011tk}.
In section 3, we explain setup of our numerical lattice QCD simulations.
In section 4, we present the resulting baryon-baryon potentials.
In section 5, we show lattice QCD implication for the $NN$ scattering and the $H$-dibaryon. 
In section 6, we estimate the flavor $SU(3)$ breaking effect on the $H$-dibaryon phenomenologically.
Section 7 is devoted to summary.
In Appendix, we consider few-body nucleon systems by using the potential obtained in section 4.

\section{Formalism}
In Refs.~\cite{Ishii:2006ec,Aoki:2009ji},  
the Nambu-Bethe-Salpeter (NBS) wave function for the two-baryon system,
\begin{equation}
 \phi_E(\vec r)e^{-Et} 
   = \sum_{\vec x} \langle 0 \vert B_i(\vec x + \vec r,t)B_j(\vec x,t) \vert B=2, E \rangle,
\label{eqn:NBS}
\end{equation}
where $B_i$ is a baryon operator and subscript $i$ distinguish octet-baryons,
is utilized to define a non-local potential
\begin{eqnarray}
\left[\frac{\nabla^2}{2\mu}-\frac{k^2}{2\mu} \right]\phi_E(\vec r) &=& \int d^3 s\, U(\vec r,\vec s) \phi_E(\vec s),
\label{eqn:potential}
\end{eqnarray}
where $E=\sqrt{k^2+M_1^2}+\sqrt{k^2+M_2^2}$ is the total energy in the center of mass system for two baryons
with masses $M_{1,2}$ and $\mu$ is the reduced mass.
An important point is that a non-local potential $U(\vec r,\vec s)$ here is independent of the energy $E$
as long as $E$ is below the meson production  threshold $E_{\rm th}$.
(For a generalization of the present formalism beyond the inelastic threshold, see ref.~\cite{Aoki:2011gt}.)

In lattice QCD simulations, we calculate the correlation function of two baryons, defined by
\begin{equation}
 \Psi(\vec r,t)
   = \sum_{\vec x} \langle 0 \vert B_i(\vec x + \vec r,t)B_j(\vec x,t) \overline{B_k} \overline{B_l} (t=0)\vert 0 \rangle,
\label{eqn:psi}
\end{equation}
which is expressed in terms of the NBS wave function as
\begin{equation}
 \Psi(\vec r, t)
  \, = \, A_{\rm gr} \phi_{E_{\rm gr}}(\vec r)e^{-E_{\rm gr}\,t} ~ + ~ 
          A_{\rm 1st}\phi_{E_{\rm 1st}}(\vec r) e^{-E_{\rm 1st}\,t} ~ \cdots
\label{eqn:tdepNBS}
\end{equation}
where $\overline{B_k} \overline{B_l} (t=0)$ is a wall source operator for two baryons at $t=0$,
defined with a baryon wall source operator
e.g. $\bar p=\sum_{x,y,z} \varepsilon_{a,b,c} ({\bar u}^{a\,T}(x)C \gamma_5{\bar d}^b(y)){\bar u}^c(z)$ for proton,
and $E_{\rm gr}$ ($E_{\rm 1st}$) is an energy of the ground(first excited) state, and $A_{\rm gr}$ ($A_{\rm 1st}$) is the corresponding coefficient.
Hereafter we call $\Psi(\vec r, t)$ a time($t$)-dependent NBS wave function. 
At large $t$ such that $(E_{\rm 1st}-E_{\rm gr}) t \gg 1$, $\Psi(\vec r, t)$ converges to  $\phi_{E_{\rm gr}}(\vec r)$
up to the ($t$-dependent) overall normalization.
In the previous study~\cite{Ishii:2006ec,Aoki:2009ji,Nemura:2008sp,Inoue:2010hs},
potentials have been successfully extracted from eq.~(\ref{eqn:potential}) by the replacement that
$\phi_E(\vec r)\rightarrow \Psi(\vec r, t)$ at large $t$, 
together with the velocity expansion of $U(\vec r,\vec s) = V(\vec r)\delta^{3}(\vec r-\vec s)+\cdots$ at the leading order,
since the spatial volume is not large so that  $(E_{\rm 1st}-E_{\rm gr}) t > 1$ can be satisfied at moderate values of $t$.

The extraction of $\phi_{E_{\rm gr}}(\vec r)$ from  $\Psi(\vec r, t)$, however,
becomes more and more difficult for larger volumes, since $E_{\rm 1st}-E_{\rm gr}$ is getting smaller
and the $t$ we need becomes larger~\cite{Inoue:2011tk}.
To overcome this practical difficulty for large volumes, we have recently proposed an 
extended method to extract the potential~\cite{Inoue:2010es,Inoue:2011tk,Ishii:2011}.
This uses the fact that $\Psi(\vec r,t)$, with the non-relativistic approximation,
satisfies a time-dependent Schr\"{o}dinger equation as
\begin{equation}
      \left[ M_1 + M_2 - \frac{\nabla^2}{2\mu} \right] \Psi(\vec r, t)
    + \int \!\! d^3 \vec r' \, U(\vec r, \vec r') \, \Psi(\vec r', t) 
    = - \frac{\partial}{\partial t} \Psi(\vec r, t) ,
 \label{eq:t-dep}
\end{equation}
at moderate values of $t$ such that $E_{\rm th} \, t \gg 1$.
(Note that relativistic kinematics can be fully taken into account 
 by introducing the second derivative in $t$ in the case $M_1=M_2$~\cite{Ishii:2011}.)
The leading term  $V(\vec r)$ of the velocity expansion is thus obtained as
\begin{equation}
  V(\vec r) = \frac{1}{2\mu}\frac{\nabla^2 \Psi(\vec r, t)}{\Psi(\vec r, t)} - 
              \frac{\frac{\partial}{\partial t} \Psi(\vec r, t)}{\Psi(\vec r, t)} - M_1 - M_2 ~.
\label{eq:vr}
\end{equation}
To ensure  numerical stability, we evaluate the $t$-derivative of $\Psi(\vec r, t)$ as 
\begin{equation}
  \left( -\frac{\partial}{\partial t} - M_1 - M_2 \right)\Psi(\vec r, t) 
\simeq \frac12 \log \left[ \frac{R(\vec r,t-1)}{R(\vec r,t+1)} \right] \Psi(\vec r, t)
\label{eq:tder}
\end{equation}
where $R(\vec r, t)$ is defined by $R(\vec r, t) = \Psi(\vec r, t)/((B_1(t)B_2(t))$ with
$B_{1,2}(t)$ being the single baryon temporal correlation function.

It has been shown in refs.~\cite{Inoue:2010es,Ishii:2011,Inoue:2011tk}
that an unique $V(r)$ within statistical errors can be obtained, 
independent on the sink-time $t$ of $\Psi(\vec r, t)$
in the range of $t$ where a single baryon correlation function $B_i(t)$ is 
saturated by the ground state.
It has been also shown that, if the spatial volume is large enough to accommodate the interaction, 
$V(r)$ is independent of the lattice volume. 
Once we obtain such a volume independent potential,
we can study observables such as binding energy and scattering phase-shifts
by solving the Schr\"{o}dinger equation in the infinite volume.

Here we mention the validity of the leading-order approximation of the non-local potential $U(\vec r, \vec{r}')$ as adopted in Eq.(\ref{eq:vr}). 
In an extensive study with quenched lattice QCD simulations at $m_{\pi}\simeq 530$ MeV,
it was previously found that the leading-order central and tensor potentials is a good approximation
at least for  $T_{\rm lab} < 100$ MeV with no sign of the next-to-leading order potentials~\cite{Murano:2011nz}.
Since the lightest pseudo-scalar meson mass in the present paper is close to that in the above study,
we expect that the phase shift calculated by the  leading-order potentials to be shown later
would be valid up to  $T_{\rm lab} \sim 100$ MeV.
It is, however,  an important future problem to check this fact by the explicit evaluation
of the next-to-leading order  potentials following the method proposed in~\cite{Murano:2011nz}.

\section{Lattice QCD simulations}

For lattice QCD simulations with dynamical quarks in the flavor $SU(3)$ limit,
we have generated gauge configurations at five different values of quark masses on a $32^3 \times 32$ lattice,
employing the renormalization group improved Iwasaki gauge action \cite{Iwasaki} at $\beta=1.83$
and the non-perturbatively $O(a)$ improved Wilson quark action. 
The lattice spacing $a$ is found to be 0.121(2) fm~\cite{CPPACS-JLQCD} and hence lattice size $L$ is 3.87 fm.
Some simulation parameters are summarized in Table~\ref{tbl:lattice}.
Hadron masses measured on each ensemble, together with the quark hopping parameter $\kappa_{uds}$,
the length of thermalized trajectory $N_{\rm traj}$ and the number of configurations $N_{\rm cfg}$,
are given in  Table~\ref{tbl:mass}.

\begin{table}[t]
\caption{Lattice parameters such as
a lattice size, an inverse coupling constant $\beta$, the clover coefficient $c_{\rm sw}$,
 a lattice spacing $a$ and a physical extension $L$. See ref.~\cite{CPPACS-JLQCD} for details.}
\label{tbl:lattice}
\smallskip
\centering
 \begin{tabular}{c|c|c|c|c}
   \hline
    size             & ~~ $\beta$ ~~ & ~~ $c_{\rm sw}$ ~~ & ~ $a$ [fm] ~ & ~$L$ [fm]~  \\
   \hline 
    $32^3 \times 32$  &   1.83  &   1.761   &  0.121(2)  & 3.87  \\
   \hline
 \end{tabular}
\end{table}

\begin{table}[t]
\caption{Quark hopping parameter $\kappa_{uds}$ and corresponding hadron masses,
 $M_{\rm ps}$, $M_{\rm vec}$, $M_{\rm bar}$
 for pseudo-scalar meson, vector meson and octet-baryon, respectively.}
\label{tbl:mass}
\smallskip
\centering
 \begin{tabular}{c|c|c|c|c}
   \hline
    $\kappa_{uds}$  & ~$M_{\rm ps}$ [MeV]~ &  ~$M_{\rm vec}$ [MeV]~& ~ $M_{\rm bar}$ [MeV]~ &
   ~$N_{\rm cfg}\,/\,N_{\rm traj}$~ \\
   \hline 
     ~0.13660~ &   1170.9(7) &   1510.4(0.9) & 2274(2) & 420\,/\,4200 \\
     ~0.13710~ &   1015.2(6) &   1360.6(1.1) & 2031(2) & 360\,/\,3600 \\
     ~0.13760~ & ~\,836.5(5) &   1188.9(0.9) & 1749(1) & 480\,/\,4800 \\
     ~0.13800~ & ~\,672.3(6) &   1027.6(1.0) & 1484(2) & 360\,/\,3600 \\
     ~0.13840~ & ~\,468.6(7) & ~\,829.2(1.5) & 1161(2) & 720\,/\,3600 \\
   \hline
 \end{tabular}
\end{table}

On each gauge configuration,
baryon two-point and four-point correlation functions are constructed from
quark propagators for the wall source with the Dirichlet boundary condition in the temporal direction.
Baryon operators at source are combined to generate the two-baryon state in a definite flavor $SU(3)$ irreducible representation,
while the local octet-baryon operators are used at sink.
To enhance the signal, 16 measurements are made for each configuration,
together with the average over forward and backward propagations in time.
Statistical errors are estimated by the Jackknife method with bin size equal to 12 for the $\kappa_{uds}=0.13840$ and 6 for others.

\section{Result of baryon-baryon potentials}

\begin{figure}[p]
\includegraphics[width=0.49\textwidth]{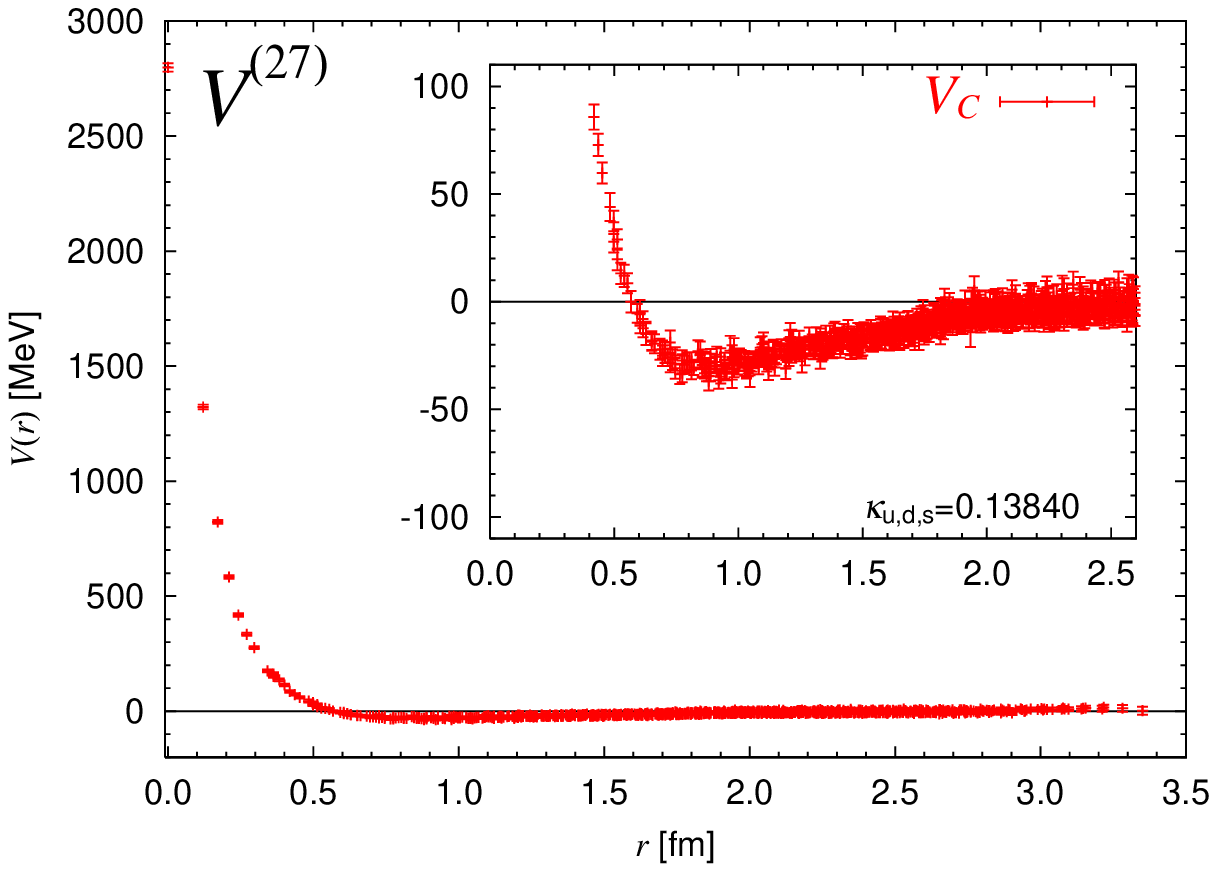} \hfill
\includegraphics[width=0.49\textwidth]{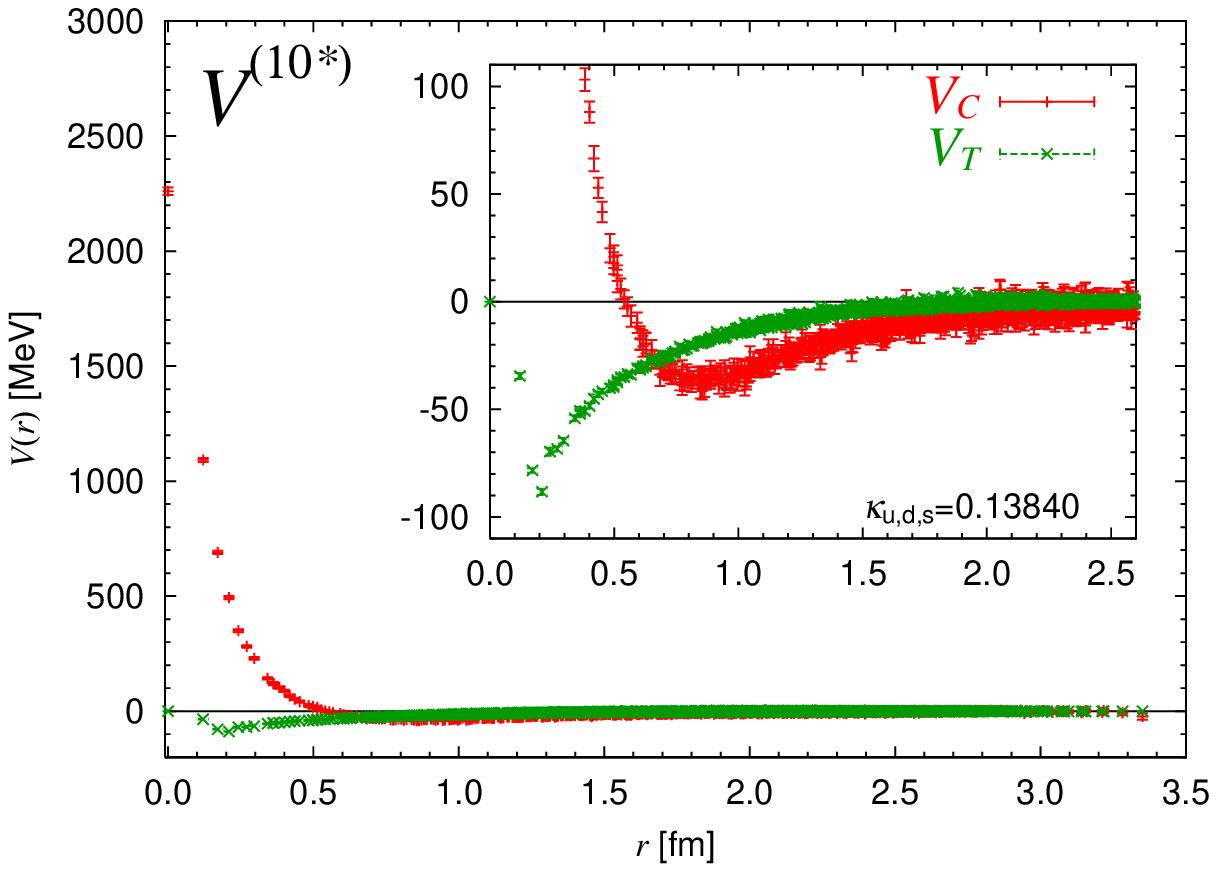}
\smallskip\newline
\includegraphics[width=0.49\textwidth]{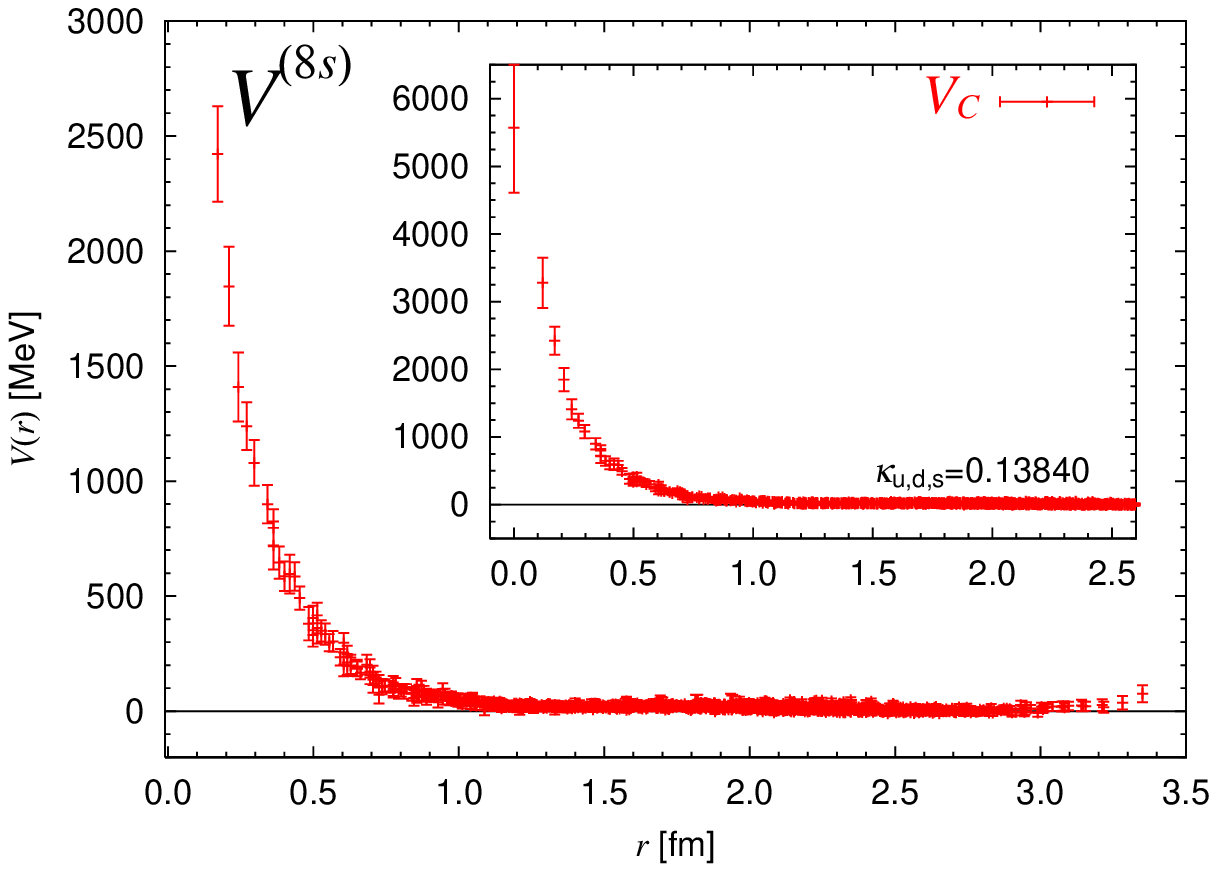} \hfill
\includegraphics[width=0.49\textwidth]{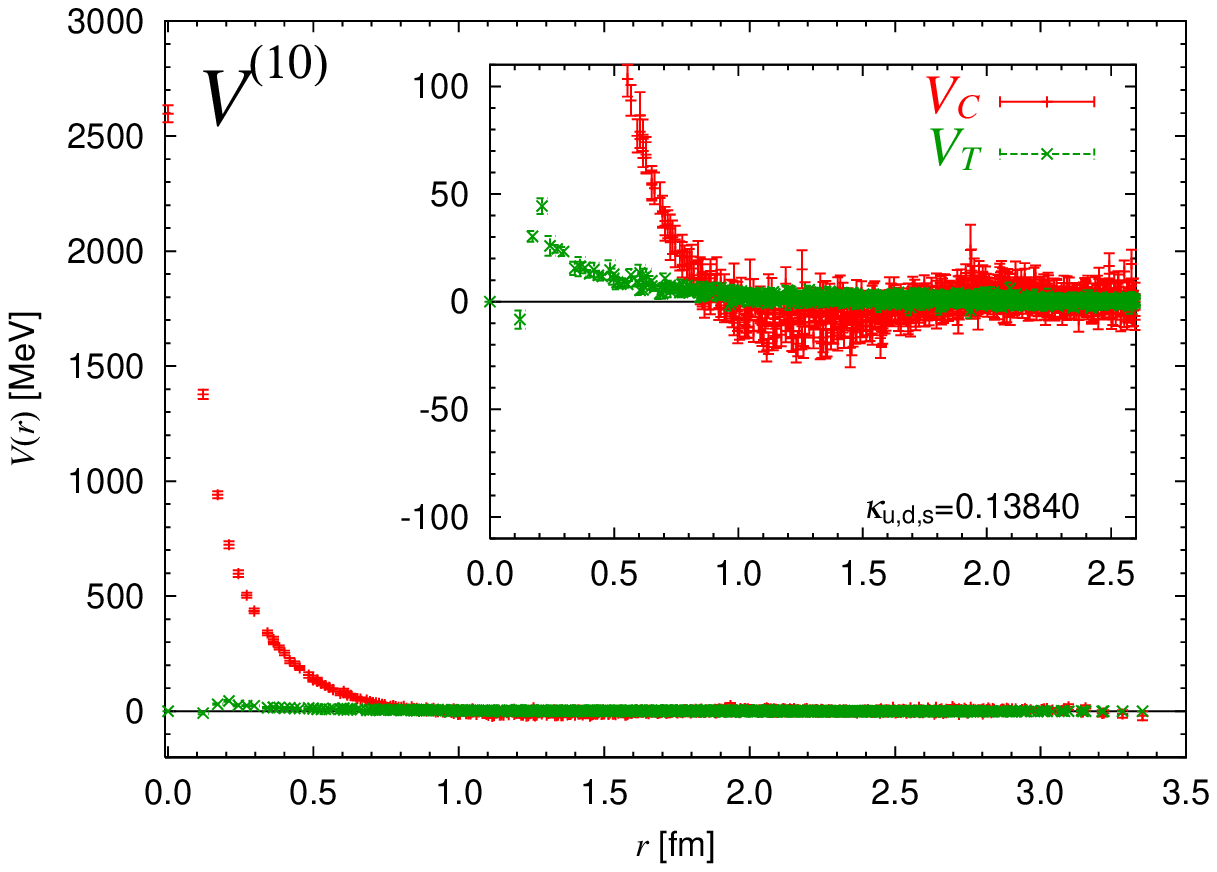}
\smallskip\newline
\includegraphics[width=0.49\textwidth]{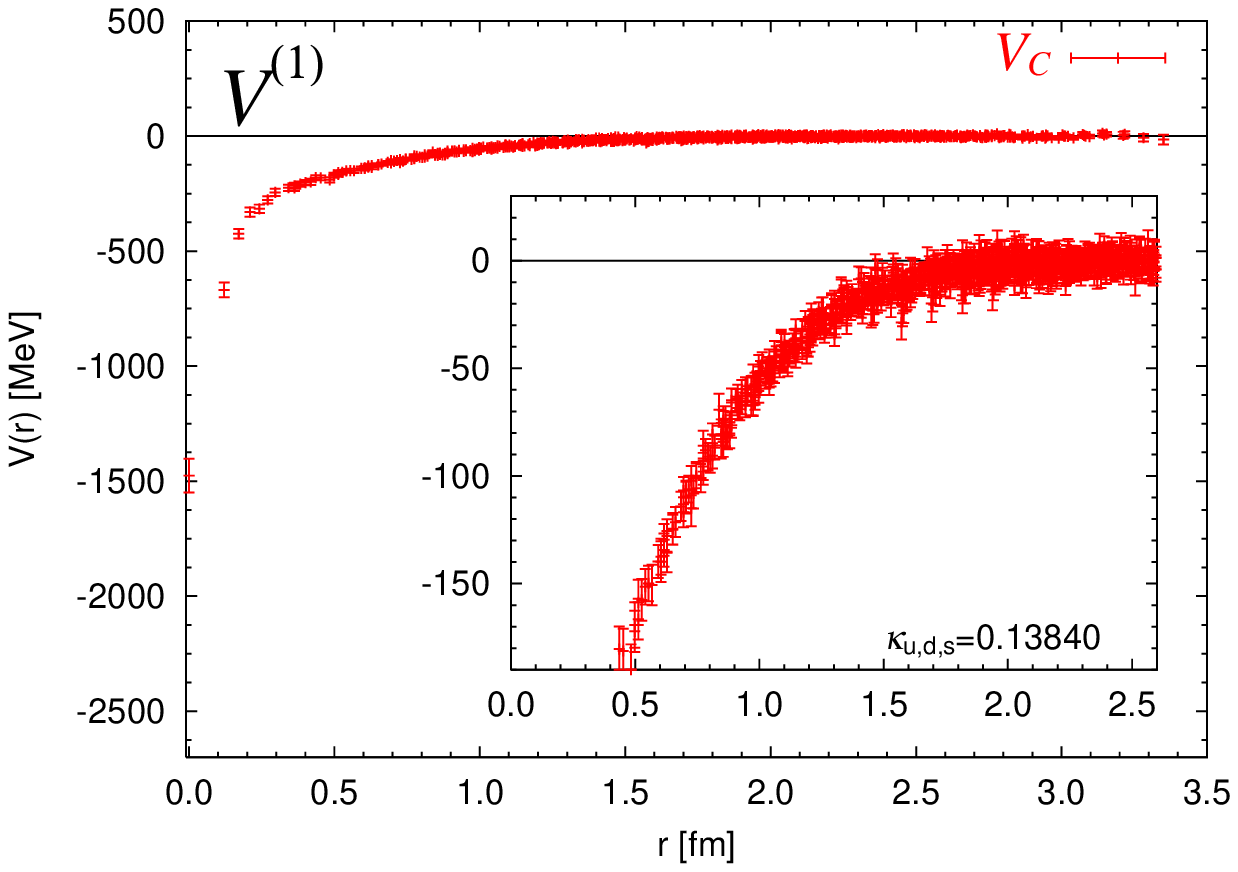} \hfill
\includegraphics[width=0.49\textwidth]{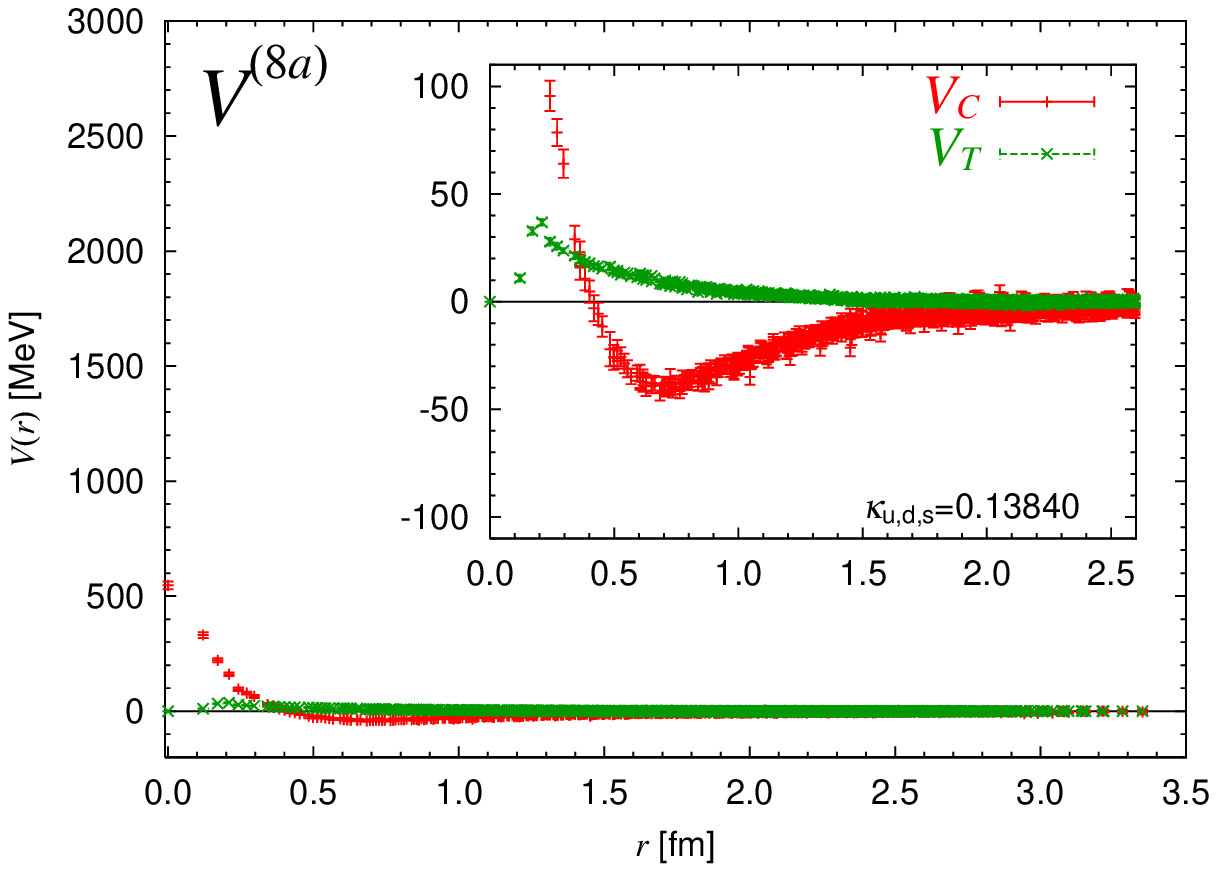}
\smallskip\newline
\caption{Potentials of baryon-baryon S-wave interaction in the flavor $SU(3)$ limit,
labeled by the flavor irreducible representation. 
These are obtained at $\kappa_{uds}=0.13840$  corresponding to the pseudo-scalar meson mass of 469 MeV.
Vertical bars indicate statistical errors estimated by the Jackknife method.
}
\label{fig:potva}
\end{figure}

In the flavor $SU(3)$ limit, all the S-wave $B\!B$ interactions are
reduced to six independent ones labeled by the flavor irreducible multiplet, 
\begin{eqnarray}
^1S_0 \ &:& \  V^{({\bf 27})}(r), \ V^{({\bf 8}_s)}(r), \ V^{({\bf 1})}(r), 
 \nonumber
\\ 
^3S_1 \ &:& \ V^{({\bf 10}^*)}(r), \ V^{({\bf 10})}(r), \ V^{({\bf 8}_a)}(r) ~.
\label{eqn:sixpot}
\end{eqnarray}
At the leading order  of the velocity expansion considered in the present report, 
we have a central potential for each channel and an additional tensor potential for each spin-triplet channel. 
Fig.~\ref{fig:potva} shows the potentials extracted from the 
data at $\kappa_{uds}=0.13840$ i.e. the lightest quark mass. 
The central potentials are qualitatively similar to ones previously obtained at heavier quark masses in a smaller volume~\cite{Inoue:2010hs}, while 
the tensor potentials for spin-triplet channels are presented for the first time in this paper.
As discussed in ref.~\cite{Inoue:2010hs},
our lattice QCD results show that the S-wave $B\!B$ interactions strongly depend on their flavor and spin.
As already discussed in ref.~\cite{Inoue:2010hs}, 
these results are consistent with the quark model prediction for S-wave $B\!B$ interactions at short distance,
indicating an important role of quark Pauli blocking effect. 

\begin{figure}[t]
\centering
\includegraphics[width=0.49\textwidth]{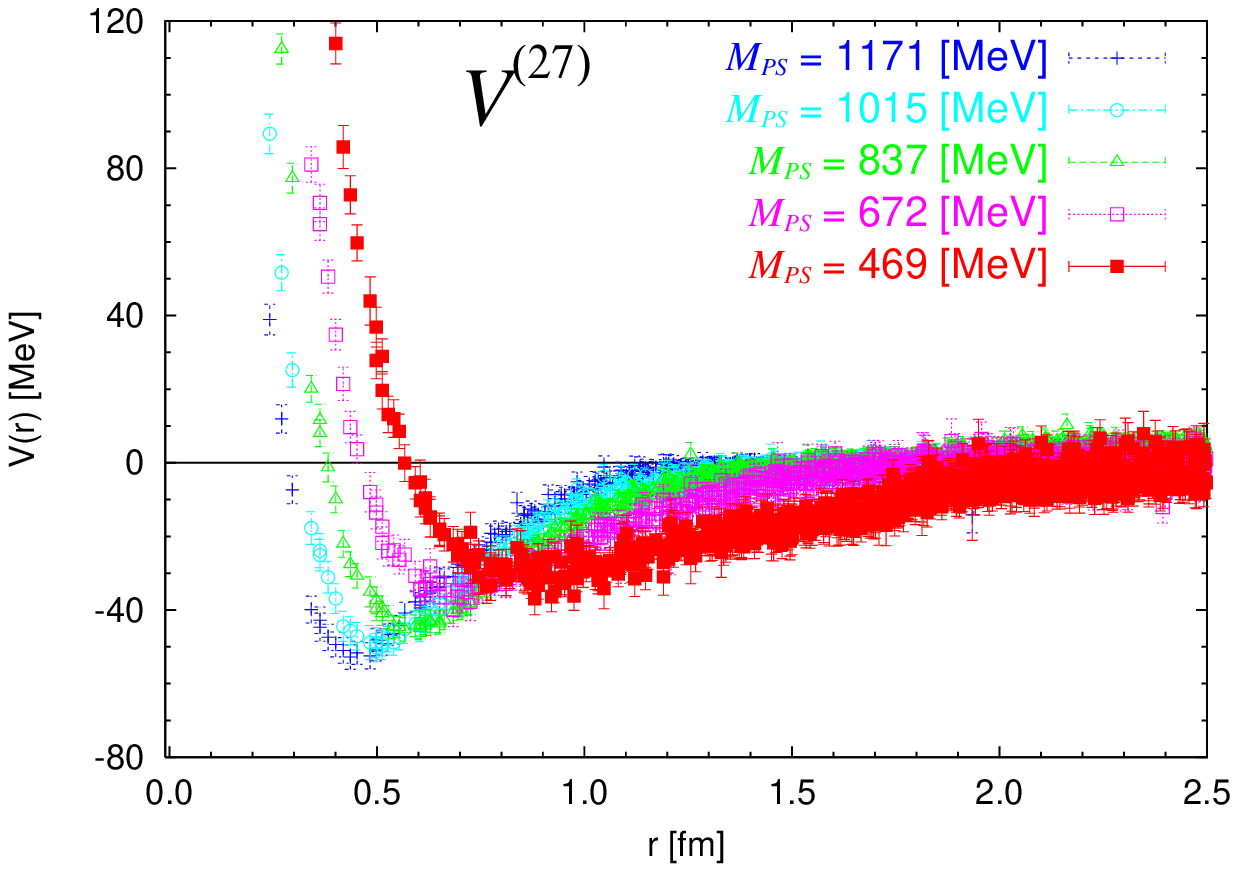}
\includegraphics[width=0.49\textwidth]{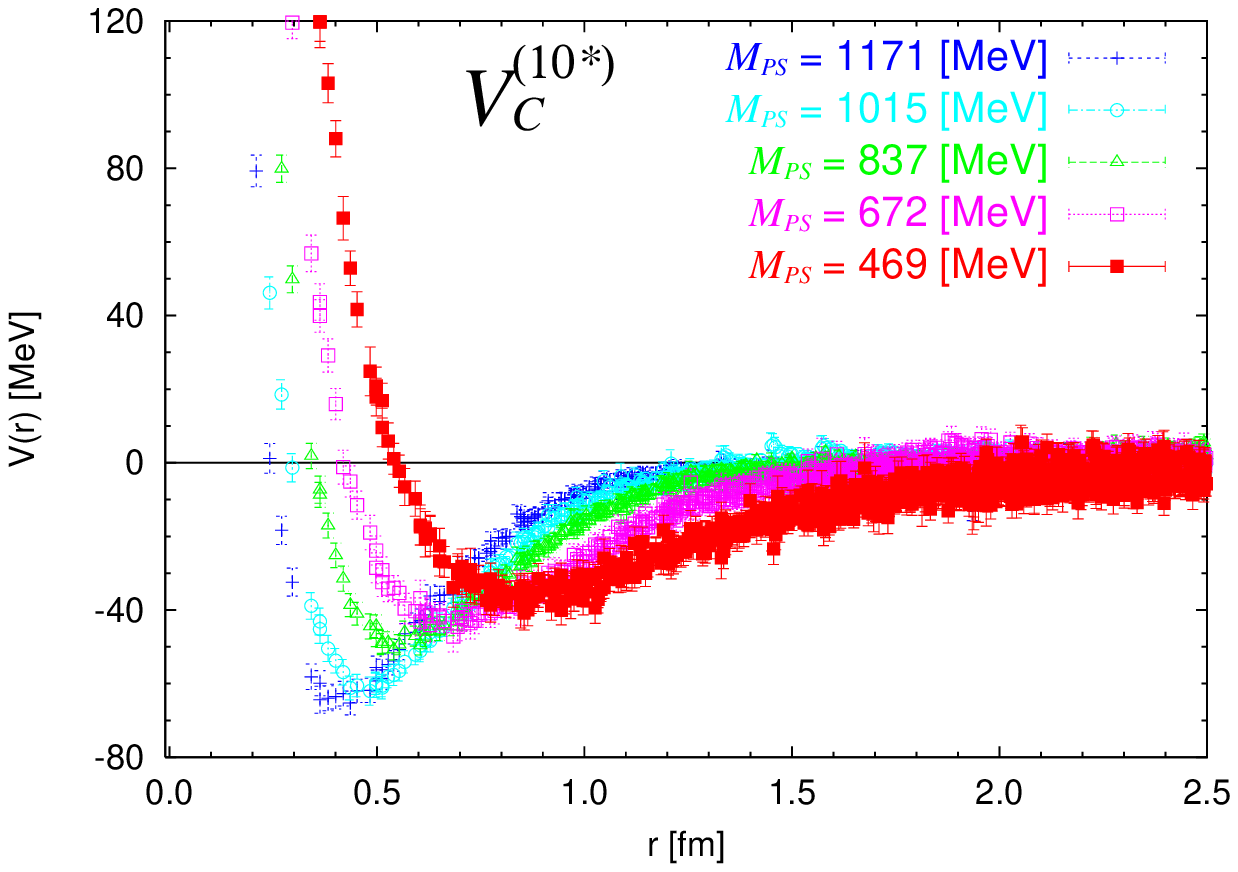}
\smallskip\newline
\includegraphics[width=0.49\textwidth]{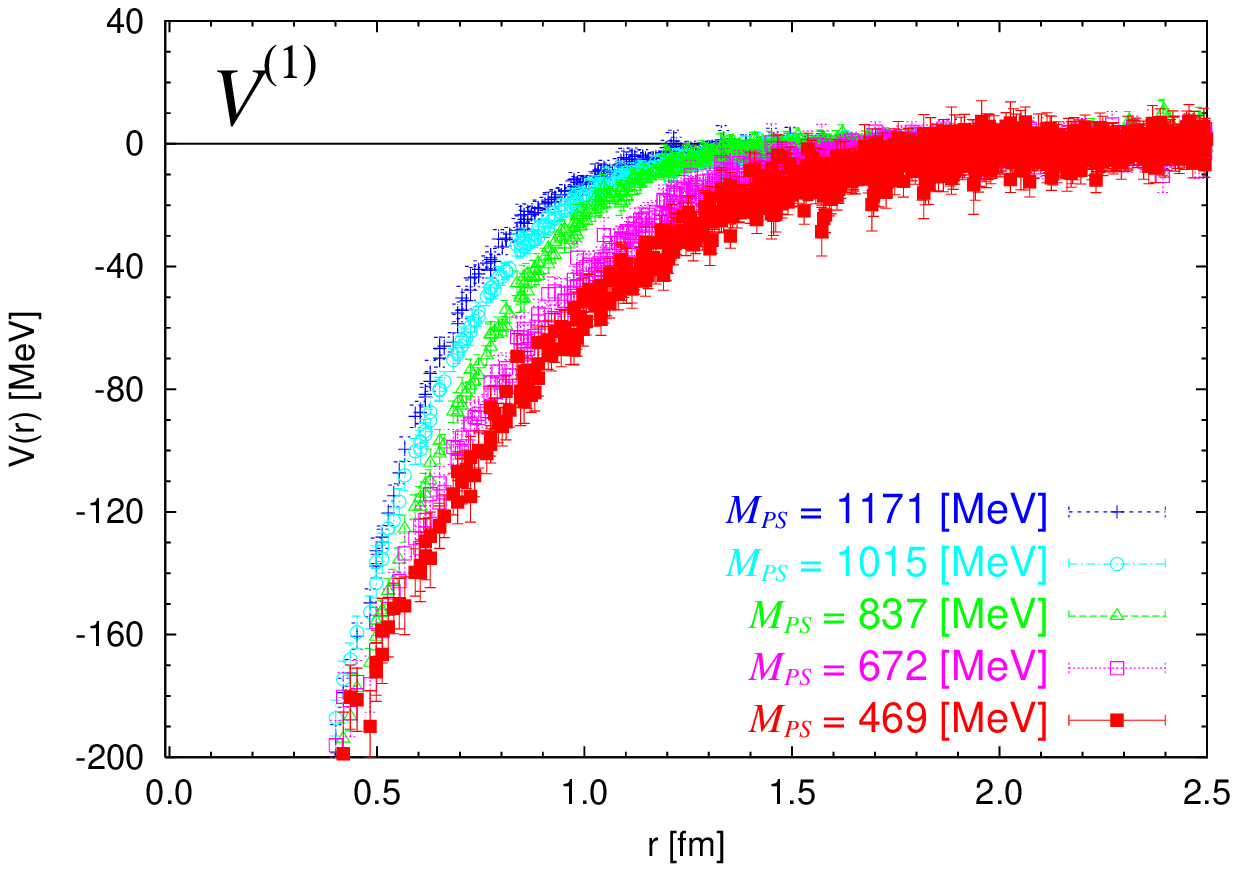}
\includegraphics[width=0.49\textwidth]{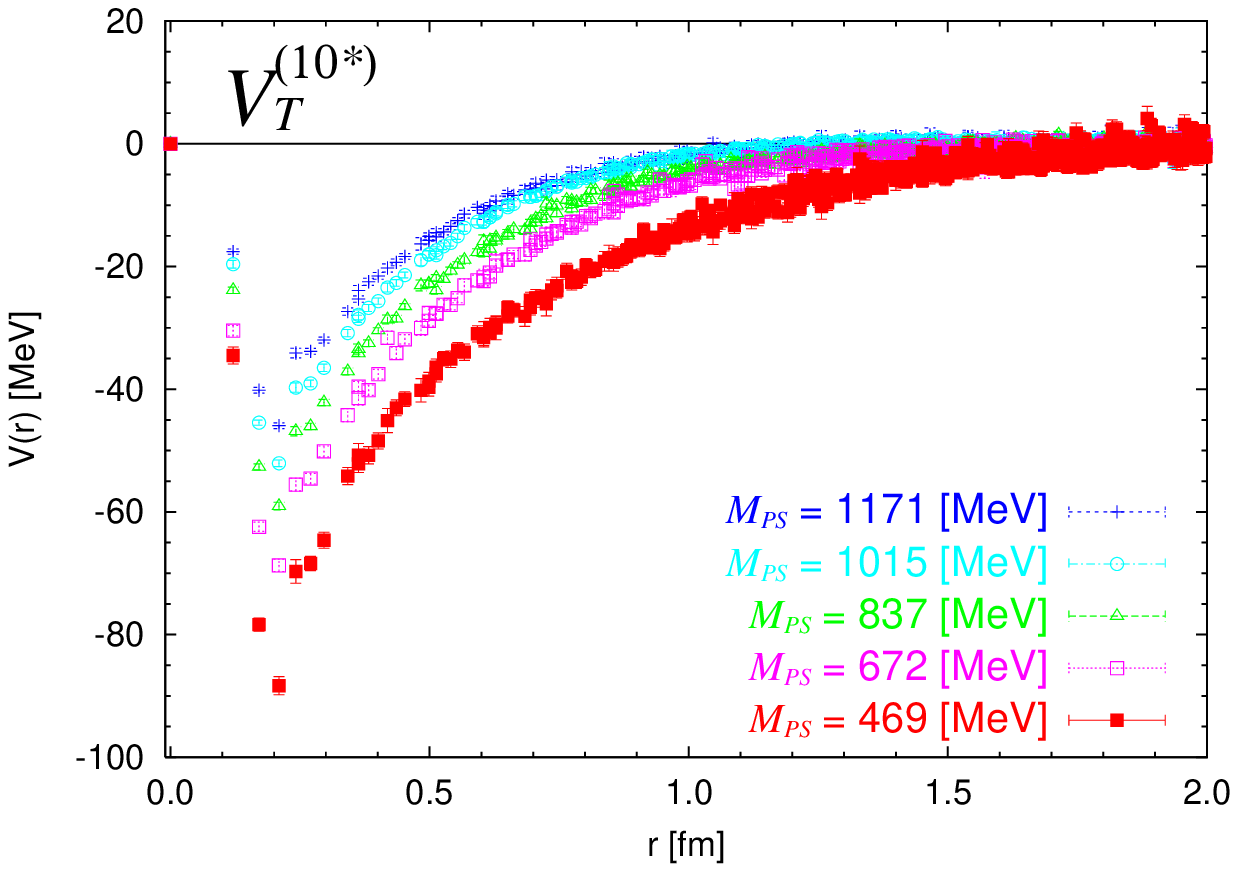}
\smallskip\newline
\caption{Quark mass dependences of  baryon-baryon potentials in the flavor $SU(3)$ limit.} 
\label{fig:pot_mass}
\end{figure}

Fig.~\ref{fig:pot_mass} shows quark mass dependences of some of these potentials.
We observe that the repulsive cores become stronger and the ranges of attractive part become longer,
in particular, the tensor force is getting stronger rapidly, as the quark mass decreases.
We therefore expect that the tensor part of $N\!N$ force becomes very strong at the physical point, $m_\pi = 135$ MeV.
This expectation is consistent with the well-known feature of the phenomenological $N\!N$ force
that the tensor force is responsible for a formation of the bound deuteron. 

\begin{figure}[t]
\includegraphics[width=0.49\textwidth]{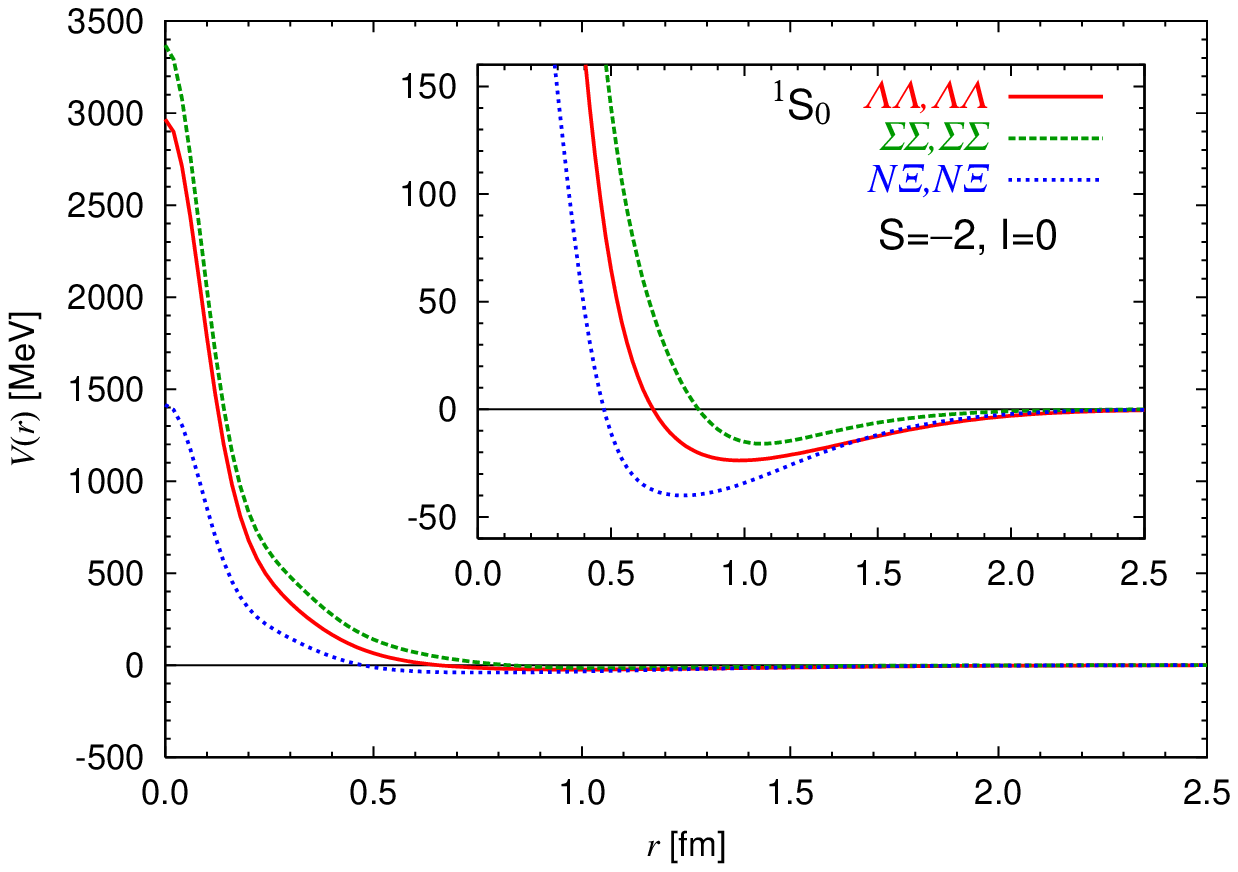}
\includegraphics[width=0.49\textwidth]{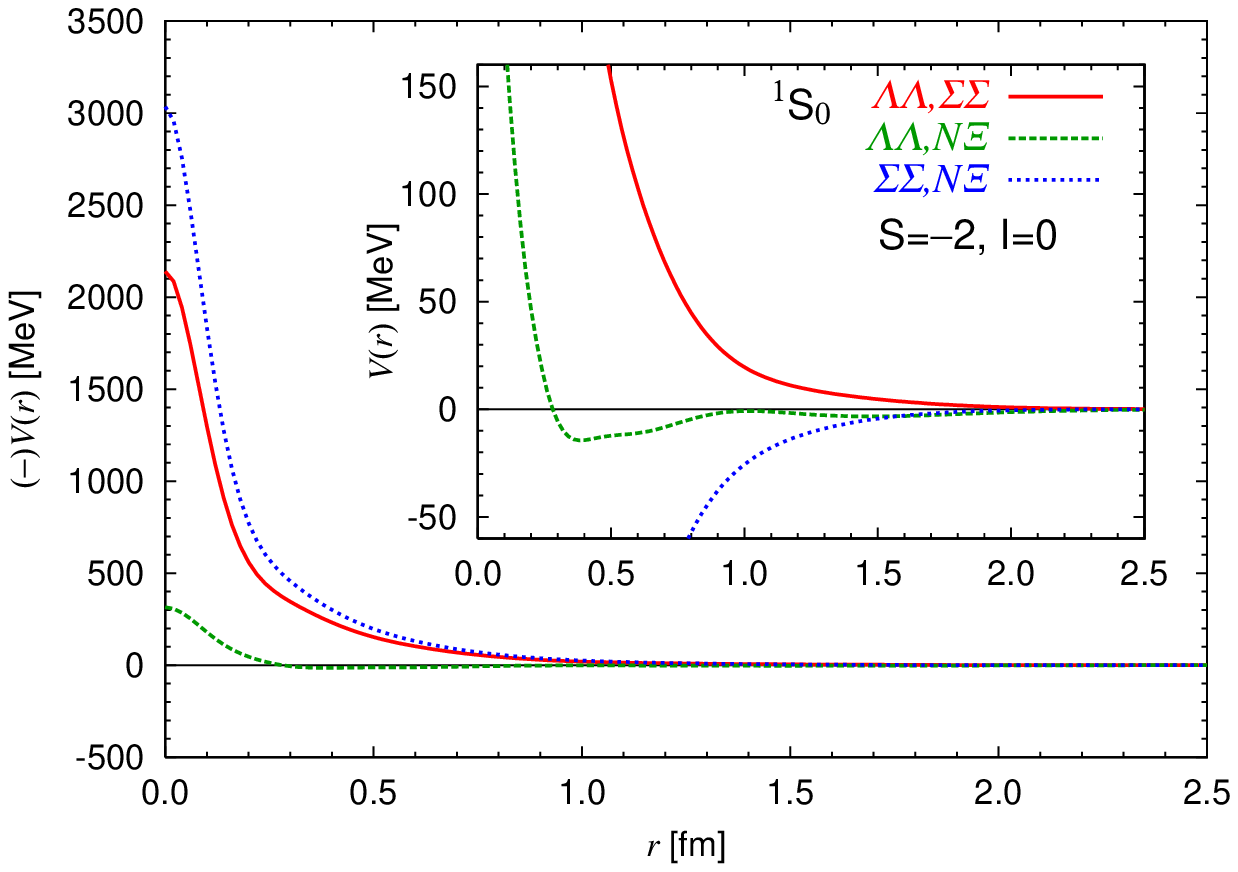}
\caption{Potentials $V_{ij}(r)$ in $S=-2$, $I=0$ sector of $^1S_0$ state,
         constructed from the present lattice QCD result and Clebsch-Gordan coefficient of $SU(3)$.}
\label{fig:potvij}
\end{figure}

In the flavor $SU(3)$ limit, the potential matrix in the particle basis, $V_{ij}(r)$, is obtained from
the potential in the flavor basis  $\mbox{diag}(V^{(1)}(r), V^{(8)}(r), V^{(27)}(r))$ by 
the unitary transformation with the Clebsch-Gordan coefficients of the $SU(3)$ group.
Fig.~\ref{fig:potvij} shows examples of $V_{ij}(r)$ in the $S=-2$ sector 
constructed from our lattice QCD result at $M_{\rm ps}=469$ MeV.
(Here we parameterized $V^{(a)}(r)$  in terms of an analytic function with seven parameters.) 
One notices that a diagonal $N\Xi$ interaction is the most attractive among three
while off diagonal $\Lambda\Lambda$-$\Sigma\Sigma$ coupling and $N\Xi$-$\Sigma\Sigma$ coupling are rather strong. 
In principle, we can extract $V_{ij}(r)$ with $SU(3)$ breaking by the coupled-channel (2+1)-flavor simulations \cite{Aoki:2011gt}. 
Preliminary studies along this direction have been already started~\cite{sasaki2010}: 
Analyses for several values of $(\kappa_{ud}, \kappa_s)$ with a larger volume will be reported elsewhere.
  
\section{Observables from $B\!B$ potentials in an infinite volume}  

\subsection{$N\!N$ phase-shifts}

\begin{figure}[t]
\includegraphics[width=0.49\textwidth]{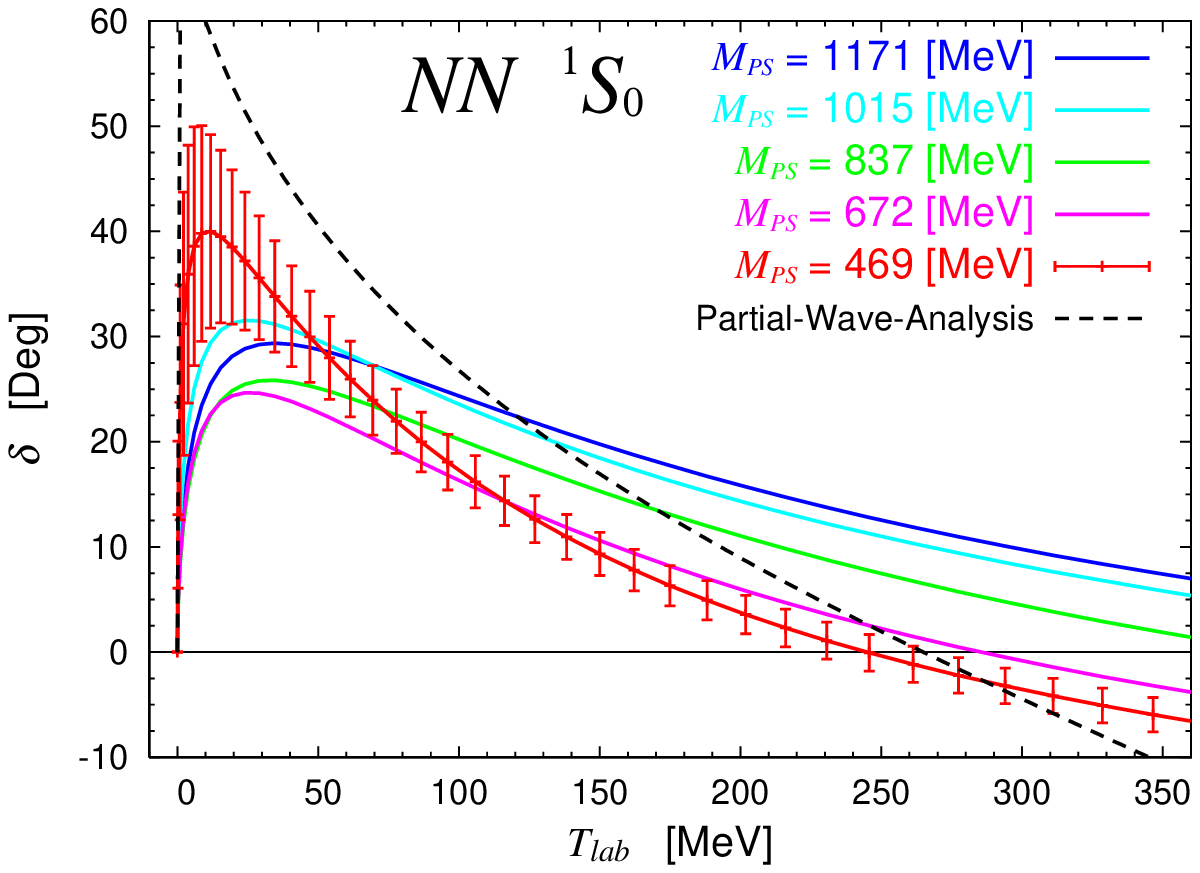}
\includegraphics[width=0.49\textwidth]{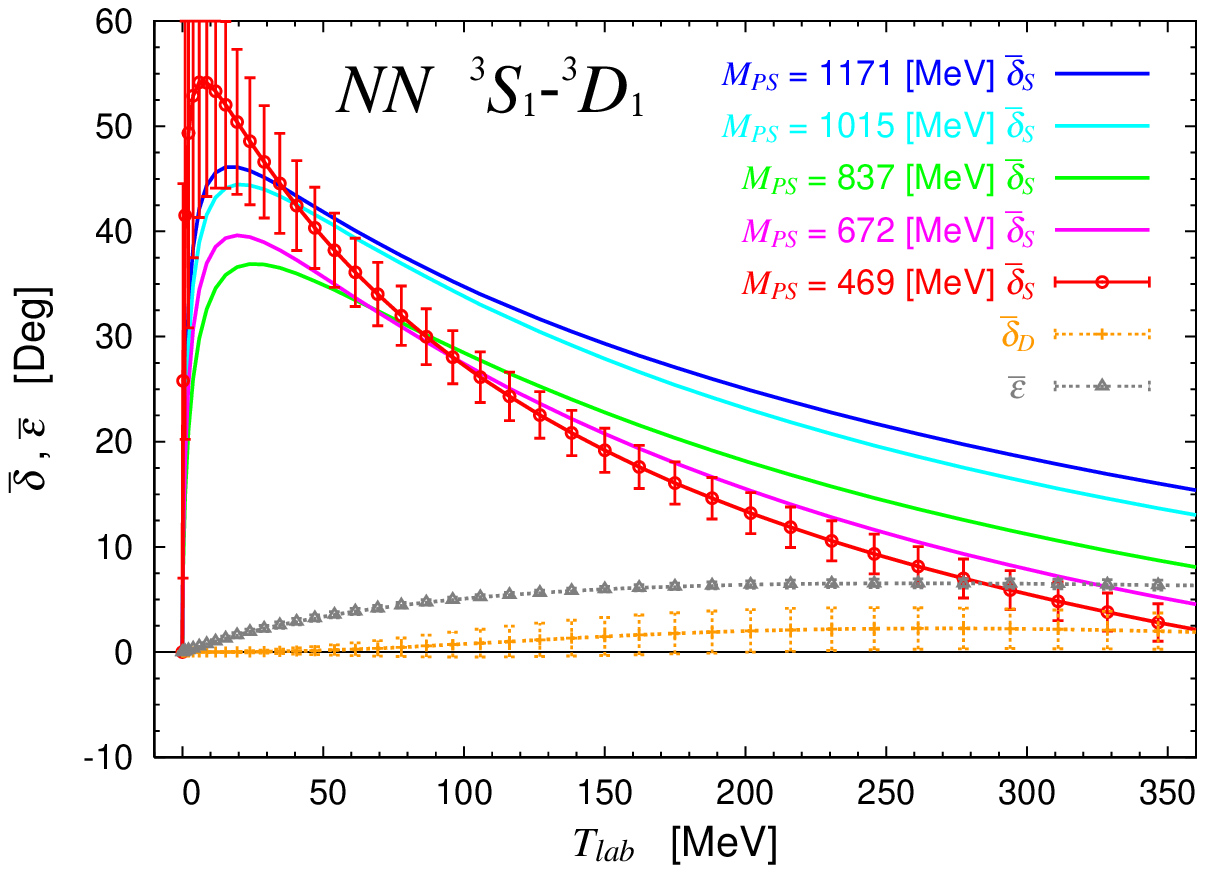}
\caption{
Phases-shifts of $N\!N$ scattering as a function of energy in the laboratory frame,
obtained from the flavor $SU(3)$ limit of lattice QCD.
In the right panel, the bar-phase-shifts and the bar-mixing-angles in the Stapp parameterization of S-matrix are plotted.
A black dashed line in the left panel represents the partial wave analysis of experimental data taken from NN-OnLine~\cite{NN-OnLine}.} 
\label{fig:nn-phase}
\end{figure}

As already explained, observables in the $B\!B$ system can be extracted from (lattice) QCD 
by solving the Schr\"{o}dinger equation involving the present potential in an infinite volume.
Fig.~\ref{fig:nn-phase} shows observables of the S-wave $N\!N$ scattering obtained in our approach,
such as the bar-phase-shifts and bar-mixing-angles as a function of energy in the laboratory frame $T_{\rm lab}$.
Due to the next-to-leading order contributions to $V(r)$ in the velocity expansion,
systematic uncertainties neglected in these figures may become sizable for  $T_{\rm lab} > 100$ MeV.
Quark mass dependences in Fig.~\ref{fig:nn-phase} reflect those of potentials in Fig.~\ref{fig:pot_mass}.
In the left panel of Fig.~\ref{fig:nn-phase}, experimental data are also plotted by a black dashed line~\cite{NN-OnLine}.
Although phase-shifts from lattice QCD depend on quark mass in a rather complicated way,
their shape approaches to the experimental one as the quark mass decreases.

As can be seen from Fig.~\ref{fig:nn-phase}, there is no bound state of two-nucleon
in both $^1S_0$ channel and $^3S_1$-$^3D_1$ channel.
We do not find any two-nucleon bound states at the wide range of $\kappa_{uds}$ corresponding $M_{p.s.}=469 - 1172$ MeV
in the $SU(3)$ limit as well as at several $(\kappa_{ud},\kappa_{s})$ combinations in 2+1 flavor QCD~\cite{Ishii:2011}.
Absence of $NN$ bound states for heavy quark masses in our simulations is in contrast to results reported in quenched QCD 
ref.~\cite{Yamazaki:2011nd} and in full QCD ref.~\cite{Beane:2011iw}
where two-nucleon bound states are found in both spin-singlet and spin-triplet channels by using the L\"{u}scher's method. 
We leave detailed comparison  among these results for  future investigations. 
In \ref{sec:fewnucl}, we consider systems with more than 2 nucleons using the present potentials:
we do not find a three-nucleon bound state, while there appears a shallow four-nucleon bound state in the case $M_{\rm ps}=469$ MeV.

\subsection{Bound $H$-dibaryon in the $SU(3)$ limit}

\begin{figure}[t]
\begin{minipage}[t]{0.475\textwidth}
\centering
\includegraphics[width=1.0\textwidth]{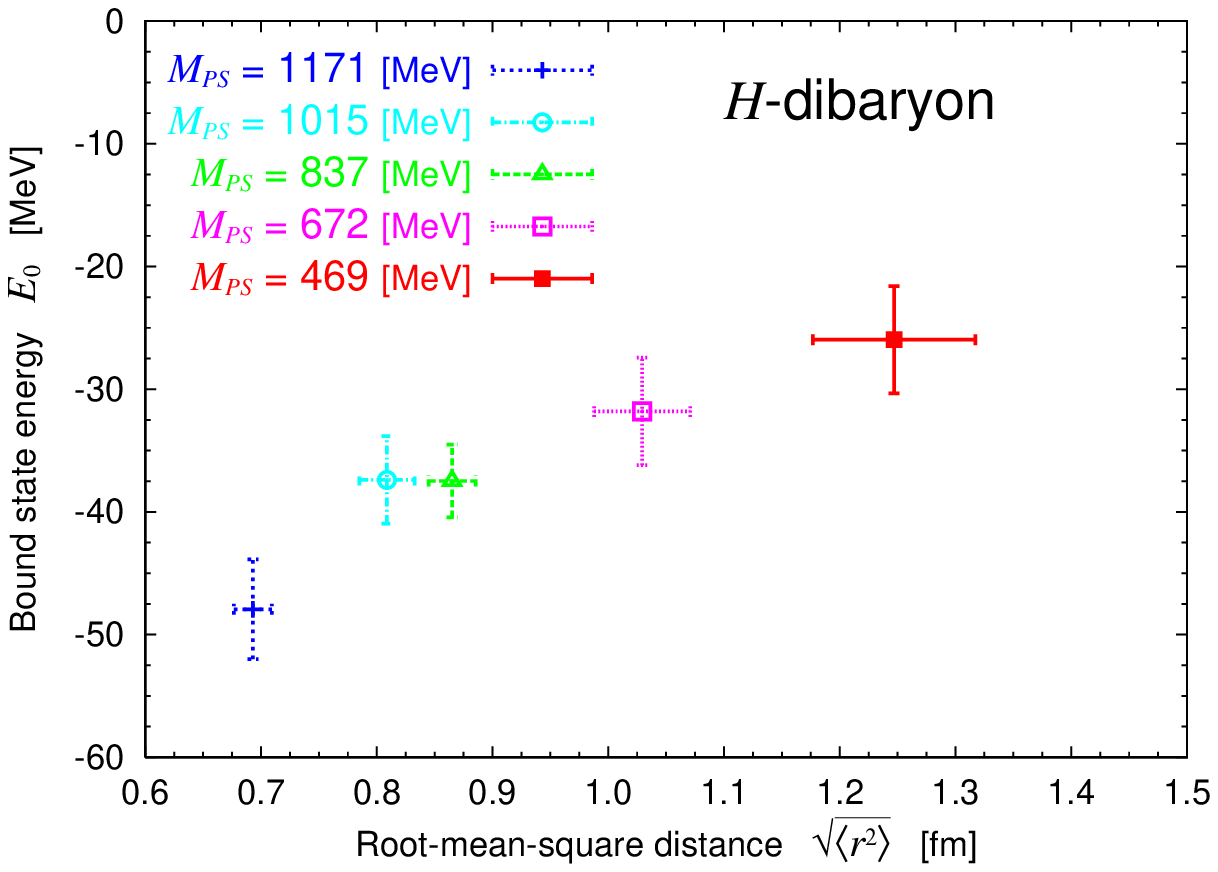}
\caption{The energy $E_0$ and the root-mean-square distance $\sqrt{\langle r^2\rangle}$ of the bound state
         in the flavor singlet channel at each quark mass.
         Bars represent statistical errors only.}
\label{fig:h1}
\end{minipage}
\hfill
\begin{minipage}[t]{0.475\textwidth}
\centering
\includegraphics[width=1.0\textwidth]{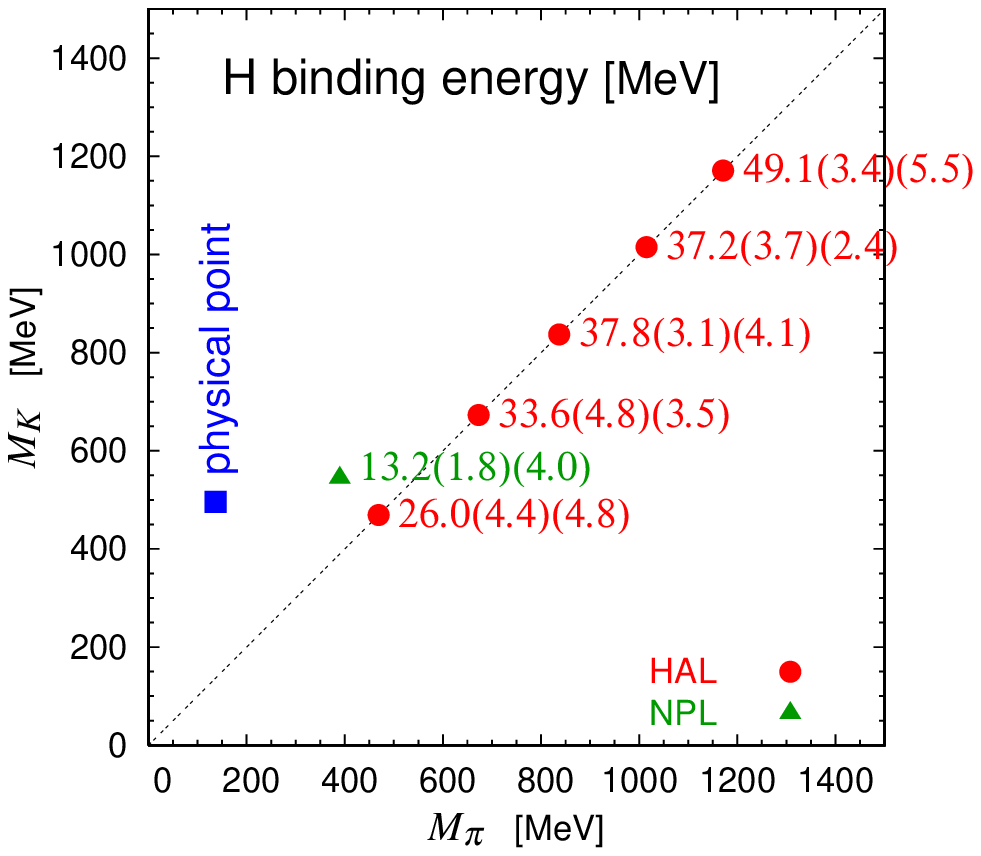}
\caption{Summary of the $H$-dibaryon binding energy in recent full QCD simulations.
 HAL stands for the present results and NPL stands for the result in ref.~\cite{Beane:2011iw}.}
\label{fig:h2}
\end{minipage}
\end{figure}

As shown in the lower left panel of Fig.~\ref{fig:potva},
the flavor-singlet potential is entirely attractive even at very short distance~\cite{Inoue:2010hs}.
By solving the Schr\"{o}dinger equation with this potential, we find a bound state in this channel~\cite{Inoue:2010es}: 
The binding energy of the $H$-dibaryon and its quark mass dependence are shown in Fig.~\ref{fig:h1},
where the energy and the size of the obtained bound state are plotted at each quark mass.
Despite the fact that the attractive potential becomes stronger as the quark mass decreases
as shown in the left lower panel of Fig.~\ref{fig:pot_mass},
the resultant binding energies of the $H$-dibaryon $\tilde B_H = -E_0$
decreases in the present range of the quark mass: This is because
the increase of the attraction toward the lighter quark mass
is compensated by the increase of the kinetic energy of  the  baryons. 
The rooted-mean square (rms) distance $\sqrt{\langle r^2 \rangle}$ is a measure of the "size" of  $H$-dibaryon,
which may be compared to the rms distance of the deuteron in nature, $1.9 \times 2 = 3.8$ fm.
Although our simulations are done for relatively heavy quark masses,
such a comparison may indicate that the $H$-dibaryon is much more compact than the deuteron.

By including a small systematic error caused by the choice of sink-time $t$ in the $t$-dependent NBS wave function,
the final result for the $H$-dibaryon binding energy becomes
\begin{eqnarray}
 M_{\rm ps}&=& 1171  ~\mbox{MeV} :~ \  {\tilde B}_H = 49.1 (3.4)(5.5) ~\mbox{MeV} \\
 M_{\rm ps}&=& 1015  ~\mbox{MeV} :~ \  {\tilde B}_H = 37.2 (3.7)(2.4) ~\mbox{MeV} \\
 M_{\rm ps}&=& ~~837 ~\mbox{MeV} :~ \  {\tilde B}_H = 37.8 (3.1)(4.2) ~\mbox{MeV} \\
 M_{\rm ps}&=& ~~672 ~\mbox{MeV} :~ \  {\tilde B}_H = 33.6 (4.8)(3.5) ~\mbox{MeV} \\
 M_{\rm ps}&=& ~~469 ~\mbox{MeV} :~ \  {\tilde B}_H = 26.0 (4.4)(4.8) ~\mbox{MeV} 
\end{eqnarray} 
with statistical error (first parenthesis) and systematic error (second parenthesis).
A bound $H$-dibaryon is also reported by the
full QCD simulation with a different approach~\cite{Beane:2010hg,Beane:2011iw}:
Their binding energy from the $\Lambda\Lambda$ threshold reads $B_H = 13.2(1.8)(4.0)$ MeV
at $(M_{\pi},M_{K})\simeq (389, 544)$ MeV, which is 
consistent with our result.
Fig.~\ref{fig:h2} gives a summary of the binding energy of the $H$-dibaryon 
obtained in recent full QCD simulations.

\section{SU(3) breaking and $H$ dibaryon}

\begin{figure}[t]
\centering
\includegraphics[width=0.49\textwidth]{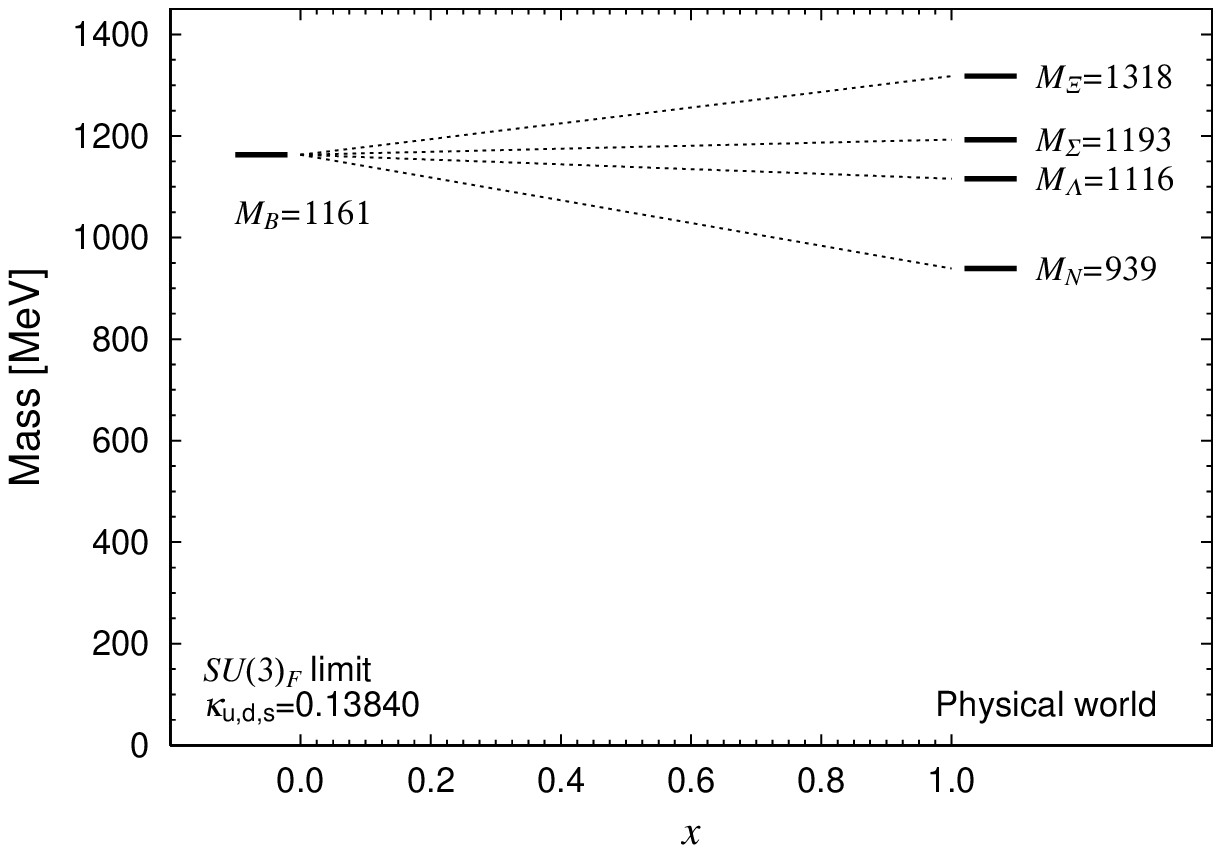}
\caption{Mass of octet baryons in the
 flavor $SU(3)$ limit at $M_{\rm ps}=469$ MeV shown at the left side($x=0$),
 and the hyperon masses in the real world shown at the right side($x=1$). 
 Dashed lines are the linear interpolation as a function of the parameter $x$.}
\label{fig:baryon_mass}
\end{figure}

When the flavor $SU(3)$ symmetry is broken, masses of octet baryons are not degenerated any more.
Fig.~\ref{fig:baryon_mass} shows masses of ``octet" baryons in the real world $M_Y^{Phys}$ plotted at the right side,
while those in the flavor $SU(3)$ symmetric world with $\kappa_{uds}=0.13840$ is plotted at the left side.
The degenerated octet baryon mass $M_Y^{SU(3)}$ is more or less equal to an average of physical ``octet'' baryon masses.
For later purpose, we introduce a phenomenological linear interpolation between the two limits,
$M_Y(x) = (1-x) M_Y^{SU(3)} + x M_Y^{Phys}$ with a parameter $x$, as shown by the dashed lines in the figure.

In broken flavor $SU(3)$ world,
the $H$-dibaryon belongs to the $S=-2$, $I=0$ sector of  $B=2$, $J^P=0^+$ states, instead of the flavor singlet channel.
There are three $B\!B$ channels in this sector i.e. $\Lambda\Lambda$, $N\Xi$ and $\Sigma\Sigma$,
which couple each other and whose interactions are described by a 3 by 3 potential matrix $V_{ij}(r)$ in the particle basis.
Observables in the real world in this sector, including the mass of the $H$-dibaryon, 
can be extracted from such a potential matrix and the baryon masses at the physical point.

Although we do not have lattice data at the physical point yet, 
let us try to make a qualitative estimate on the fate of the $H$-dibaryon in the $SU(3)$ broken world by using
the dashed lines in Fig.~\ref{fig:baryon_mass} together with the potential matrix $V_{ij}(r)$ in the flavor $SU(3)$ limit in Fig.~\ref{fig:potvij}.
This is based on the assumptions that
(i) the major effect of the $SU(3)$ breaking comes from the baryon mass splittings, 
and (ii) the qualitative features of the hyperon interactions in  Fig.\ref{fig:potvij} remain intact even with the $SU(3)$ breaking.
Validity of these assumptions should be checked in the future lattice simulations with explicit flavor $SU(3)$ breaking.
    
With the assumptions (i) and (ii), we study the spin-singlet S-wave scattering in the
coupled channel system, $\Lambda\Lambda$-$N\Xi$-$\Sigma\Sigma$, 
to trace the $H$-dibaryon as a function of the parameter $x$ in Fig.~\ref{fig:baryon_mass}.

\begin{figure}[p]
\includegraphics[width=0.49\textwidth]{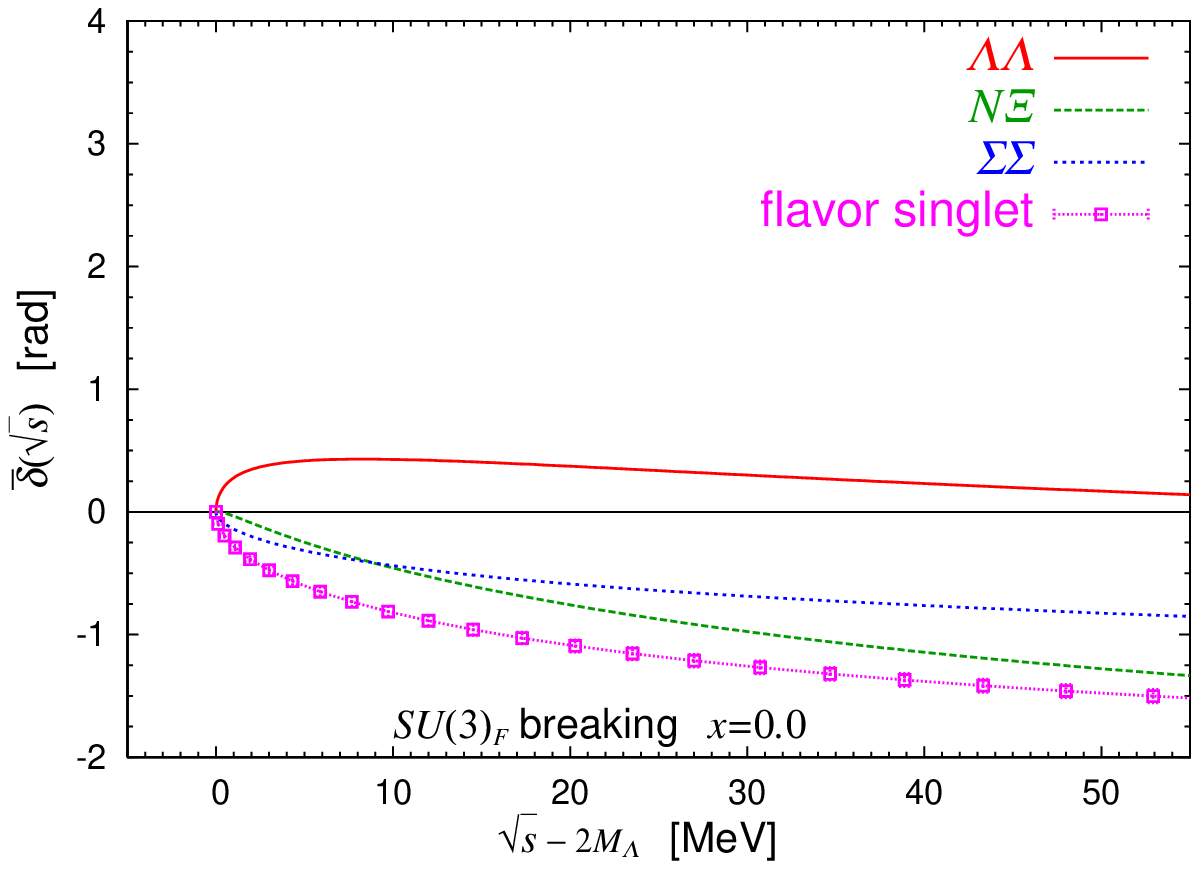} \hfill
\includegraphics[width=0.49\textwidth]{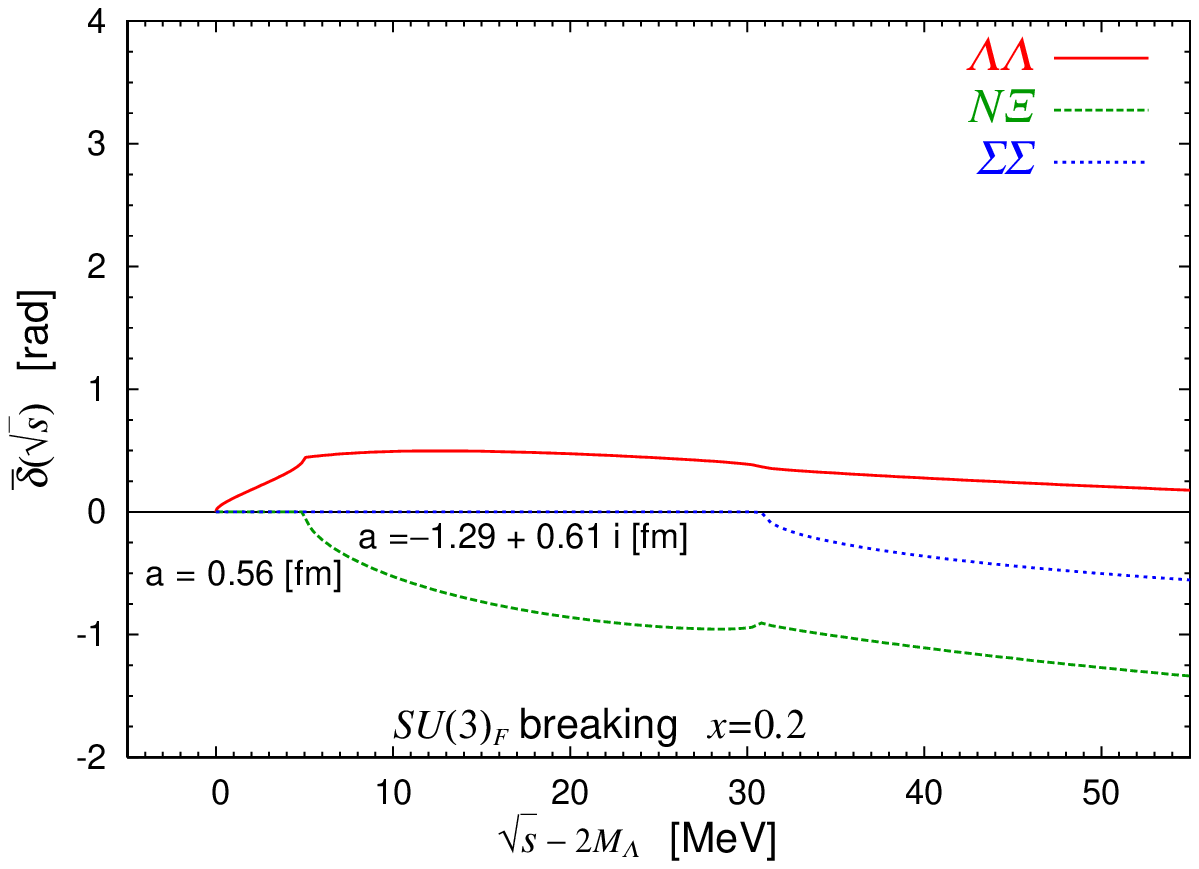}
\smallskip\newline
\includegraphics[width=0.49\textwidth]{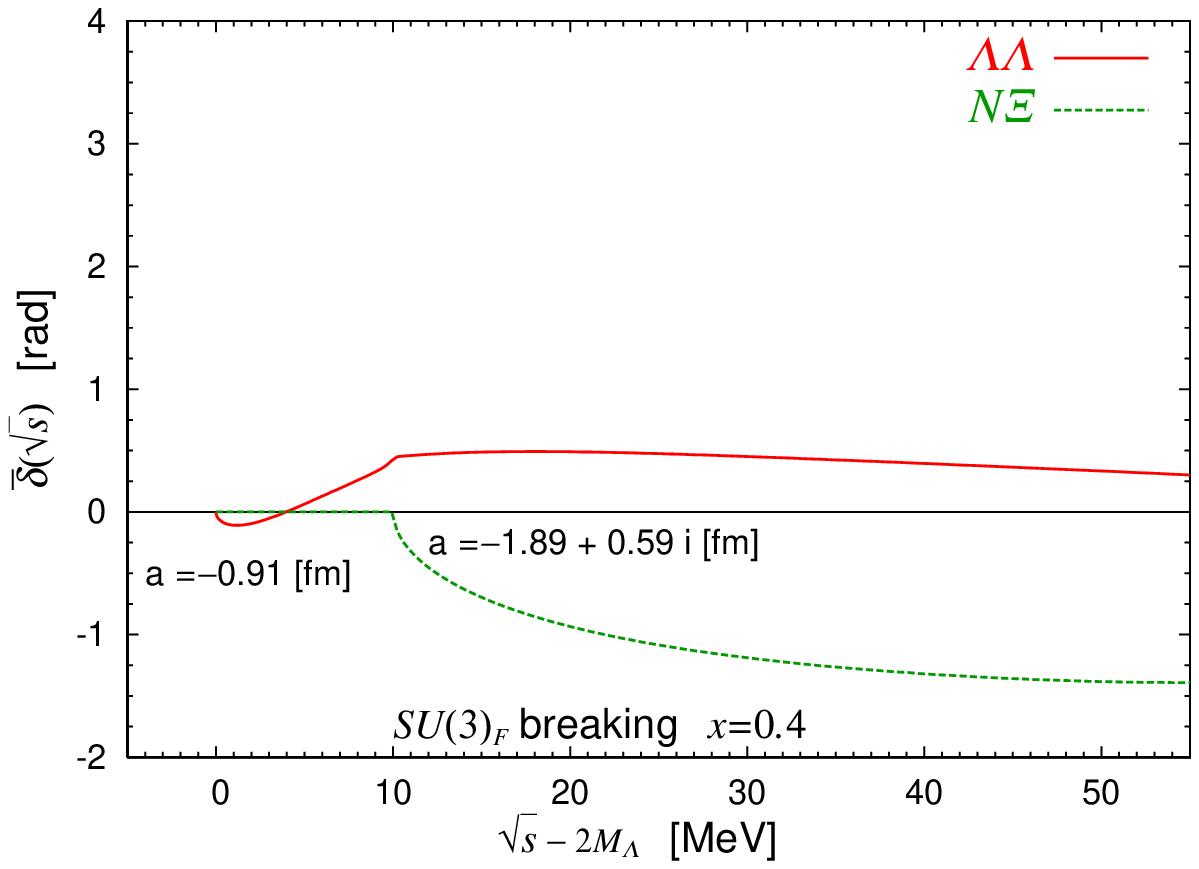} \hfill
\includegraphics[width=0.49\textwidth]{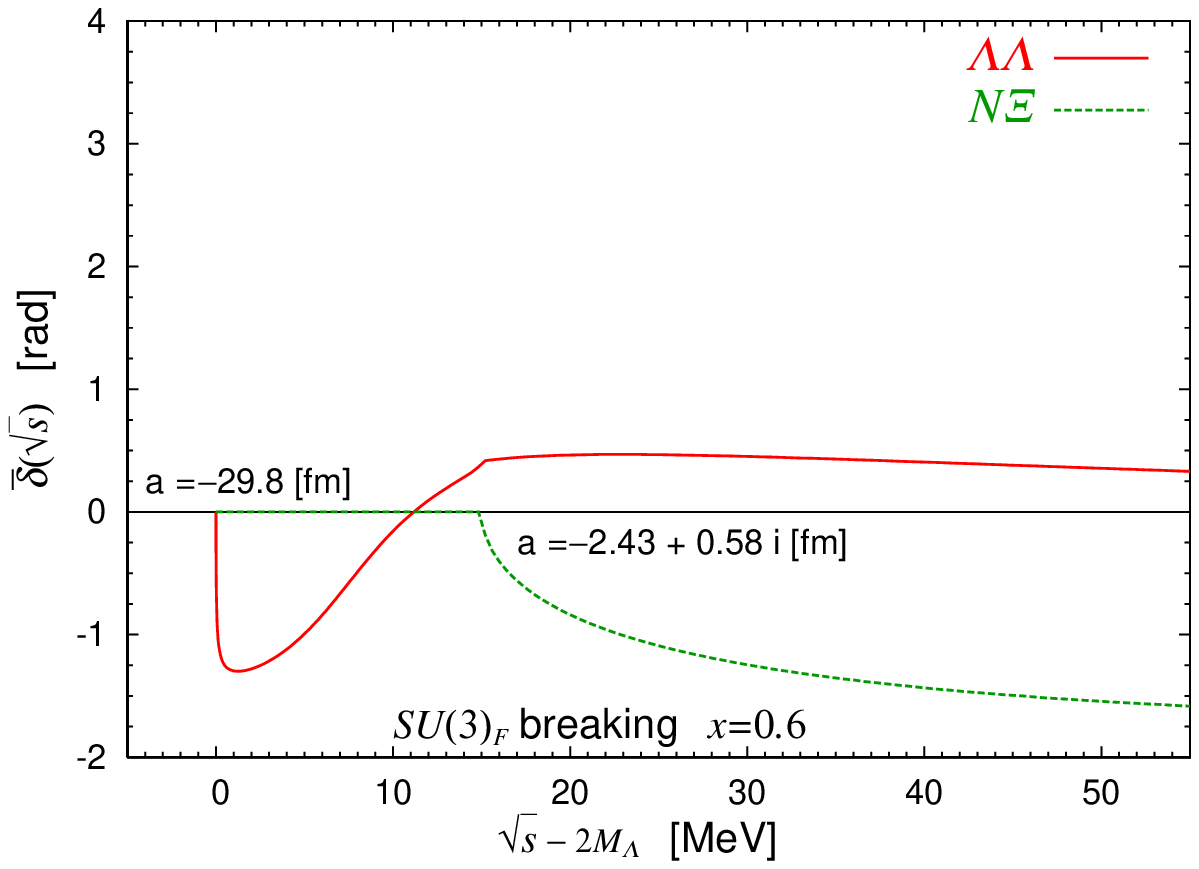}
\smallskip\newline
\includegraphics[width=0.49\textwidth]{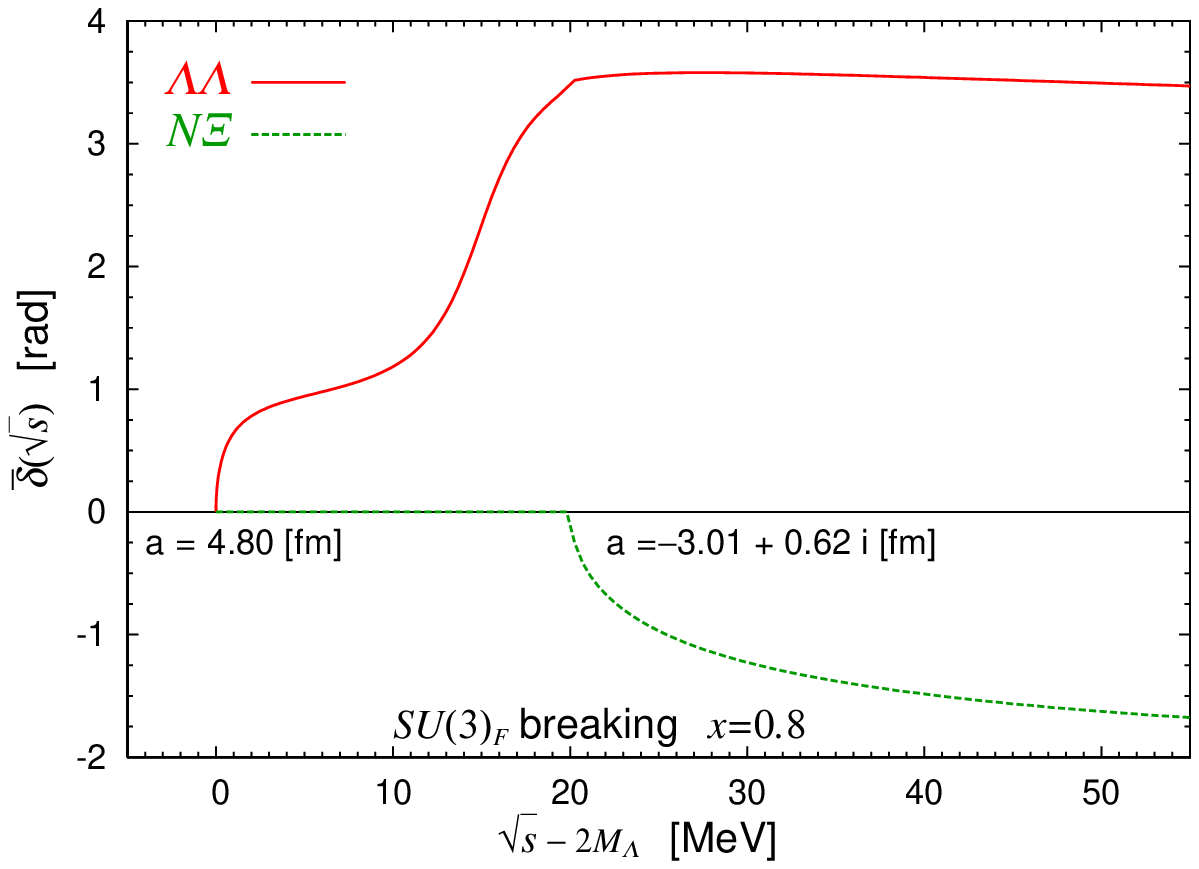} \hfill
\includegraphics[width=0.49\textwidth]{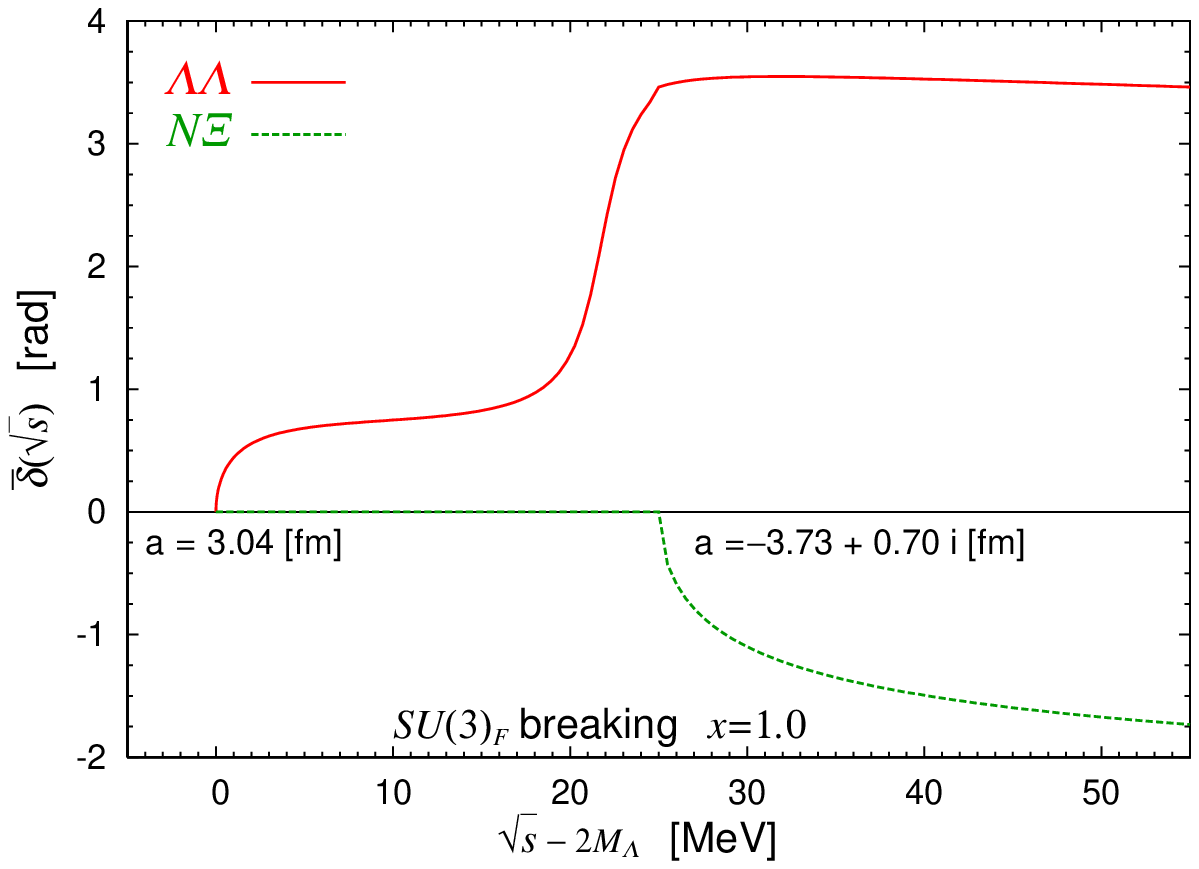}
\smallskip
\caption{The bar-phase-shifts of the baryon-baryon ${}^1S_0$ scattering in $S=-2$, $I=0$ sector,
         as a function of energy in the center of mass frame at several values of the $SU(3)$ breaking parameter $x$. 
         The scattering length $a_i$ are also shown. 
         In the top left panel for the $x=0$ case, the phase-shifts 
         in the flavor-singlet channel are also given for a reference.
         }
\label{fig:barphase}
\end{figure}

We use the bar-phase-shift $\bar{\delta}_i$ and the elasticity $\eta_i$ for the $i$-th channel,
defined by the on-energy-shell element of the S-matrix, 
$S^{l=0}_{ii} = {\eta}_i\, e^{2 i\bar{\delta}_i}$.
The scattering length $a_i$ is the amplitude at the corresponding threshold; 
for example, $a_{N\Xi} = \lim_{\sqrt s \to M_N + M_{\Xi}} (S_{N\Xi,N\Xi}^{l=0} -1 )/(2 i k)$
with the on-energy-shell momentum $k$.
To obtain the S-matrix, we solve the Lippmann-Schwinger equation for the T-matrix in momentum space, which is given by
\begin{equation}
 T^{\alpha\beta} = V^{\alpha\beta} + \sum_{\gamma}
 V^{\alpha\gamma} \, G^{(0)}_{\gamma} \, T^{\gamma\beta}, \quad 
 G^{(0)}_{\gamma} = \frac{1}{E - H^{(0)}_{\gamma} + i \epsilon},
\end{equation}
where the momentum indices are suppressed
and $H^{(0)}_{\gamma} = -{p^2}/{(2\mu_{\gamma})} + M^{\gamma}_{1} + M^{\gamma}_{2}$ is the free Hamiltonian of a channel $\gamma$.
With the analytic expression of the potentials $V_{ij}(r)$ obtained by 
fitting the lattice result,  their matrix elements in momentum space
are evaluated straightforwardly by numerical integration.   

In Fig.~\ref{fig:barphase},  $\bar{\delta}_i$ is plotted
as a function of energy in the center of mass frame at several values of the parameter $x$. 
At $x=0$, an attractive nature of the flavor singlet potential can be seen in the $\Lambda\Lambda$ phase-shift.
The sign of an existence of the H dibaryon, however, can not be clearly seen in the behaviors of the $\Lambda\Lambda$ bar-phase-shift at $x=0$,
contrary to the single channel analysis with the flavor singlet potential,
since the $H$-dibaryon is deeply bounded, 26 MeV below the  $\Lambda\Lambda$ threshold in this case.
As $x$ increases from zero, attractive $\Lambda\Lambda$ phase-shift is getting smaller and it finally becomes repulsive. 
For example, the scattering length is negative ($a=-0.91$ fm) at $x=0.4$ since the $H$ dibaryon exists below
but close enough to the $\Lambda\Lambda$ threshold.  
At $x=0.6$, the binding energy of the $H$ dibaryon becomes almost zero,
so that the scattering length becomes very large( $a=-29.8$ fm).
As $x$ further increases, the $H$ dibaryon goes above the  $\Lambda\Lambda$ threshold at a little above $x=0.6$.
In the bottom two panels of Fig.~\ref{fig:barphase}, we observe an appearance of 
the $H$ dibaryon as the resonance at $\bar\delta\simeq \pi/2$ in the case that $x=0.8$ and $1.0$.

\begin{figure}[p]
\includegraphics[width=0.49\textwidth]{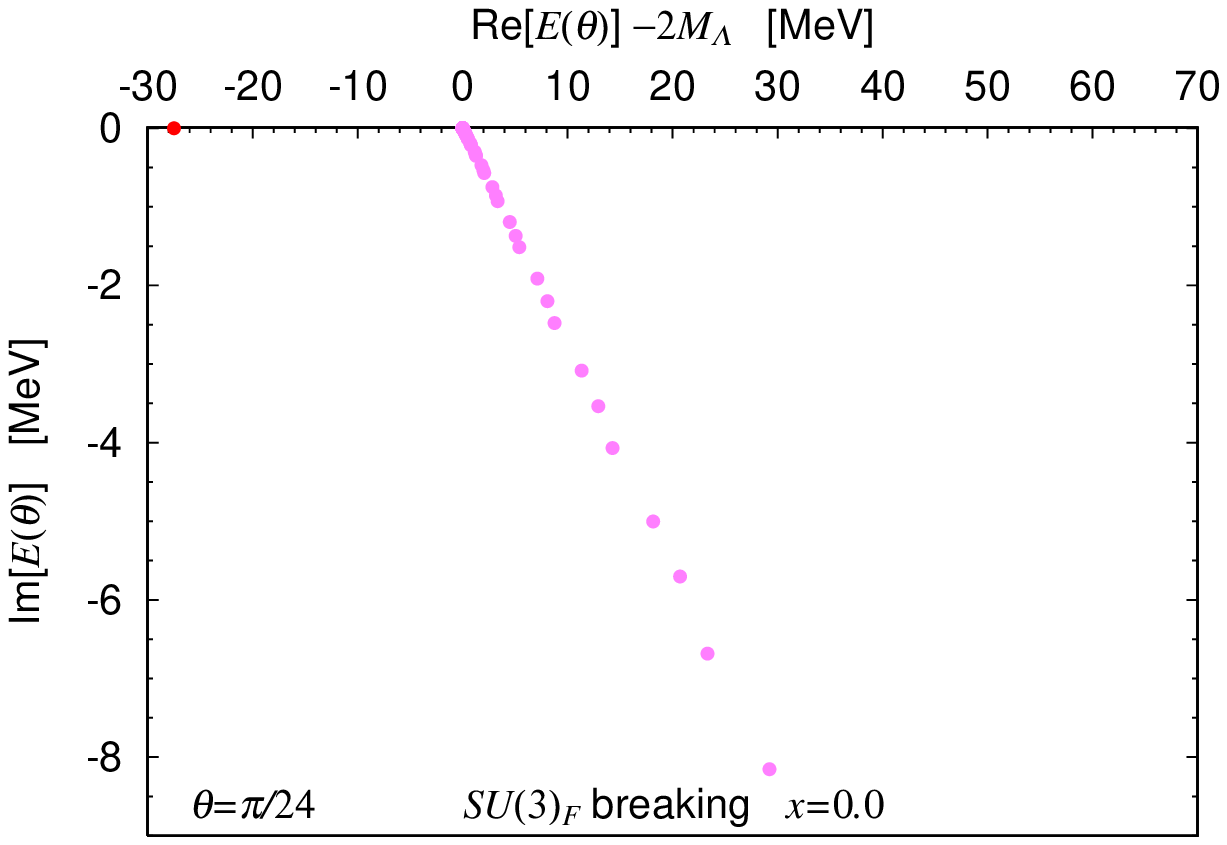} \hfill
\includegraphics[width=0.49\textwidth]{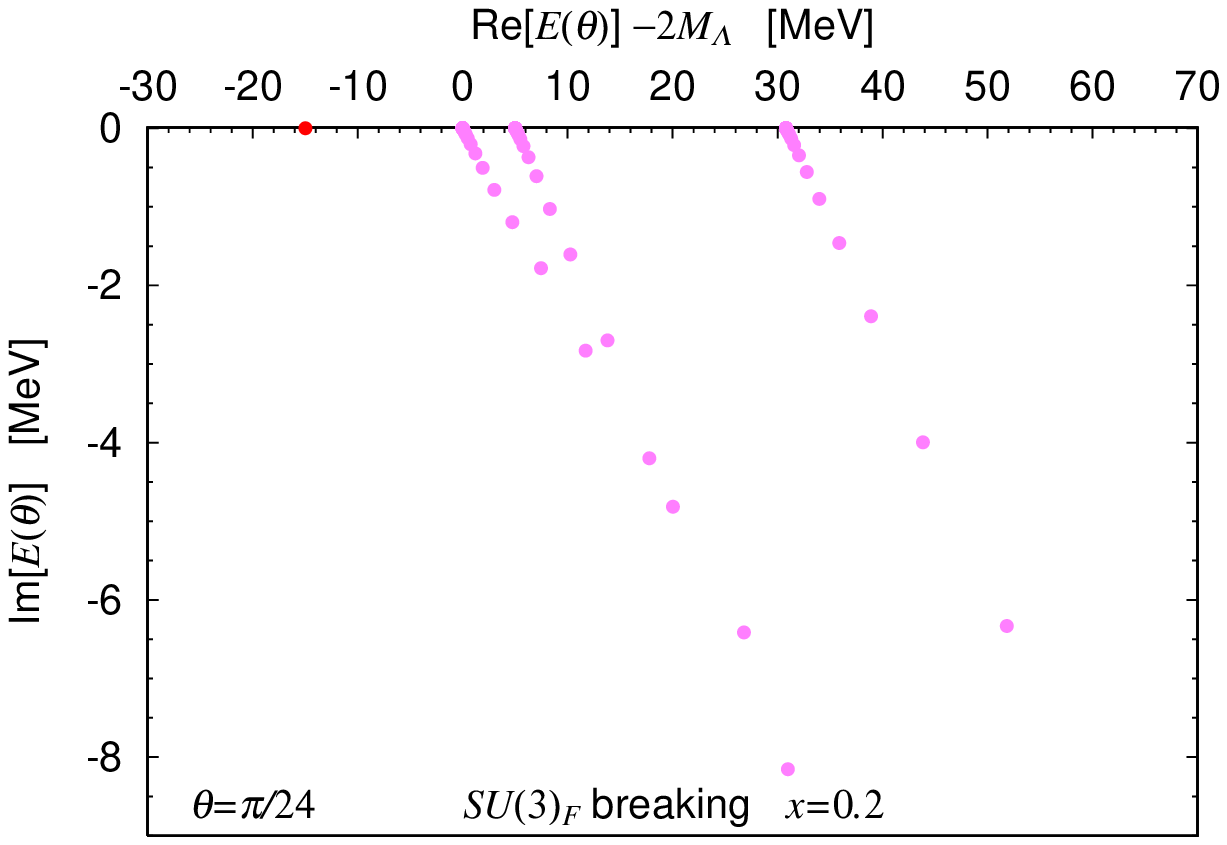}
\smallskip\newline
\includegraphics[width=0.49\textwidth]{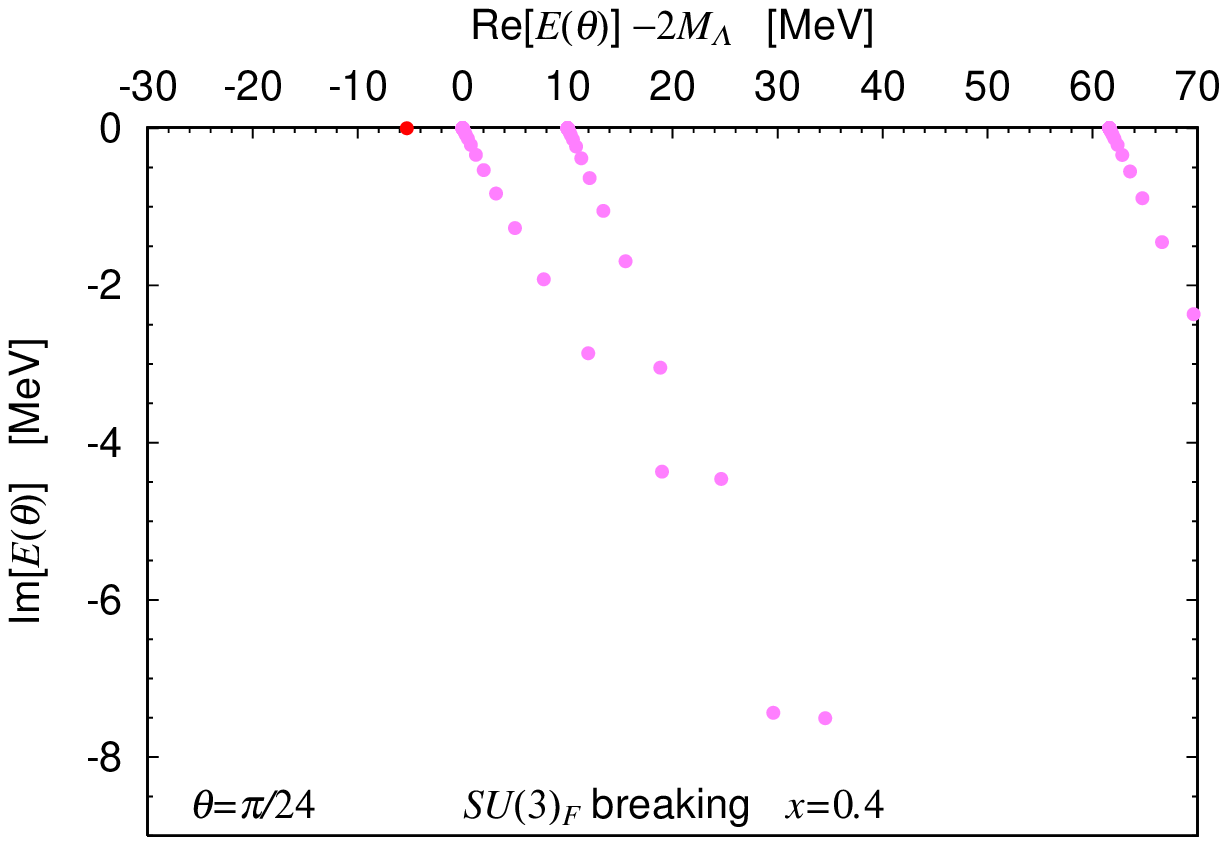} \hfill
\includegraphics[width=0.49\textwidth]{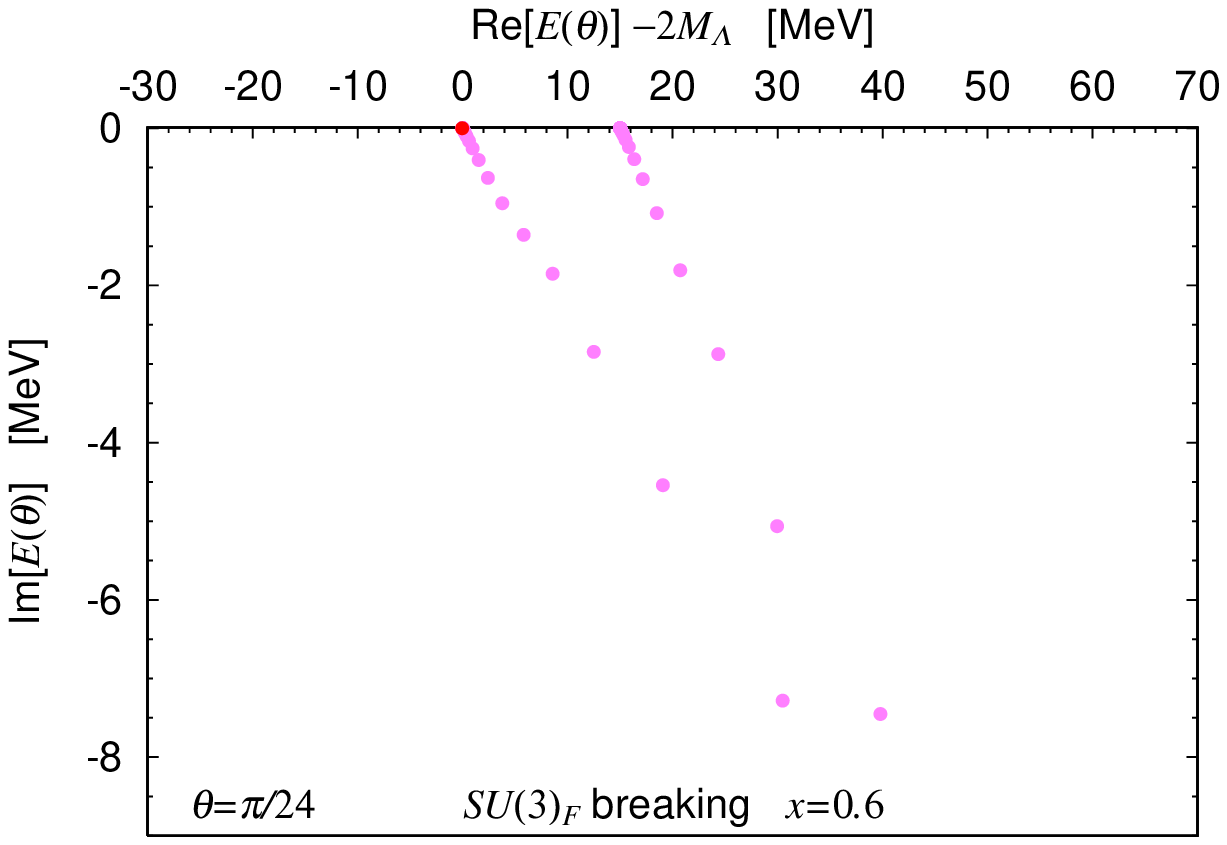}
\smallskip\newline
\includegraphics[width=0.49\textwidth]{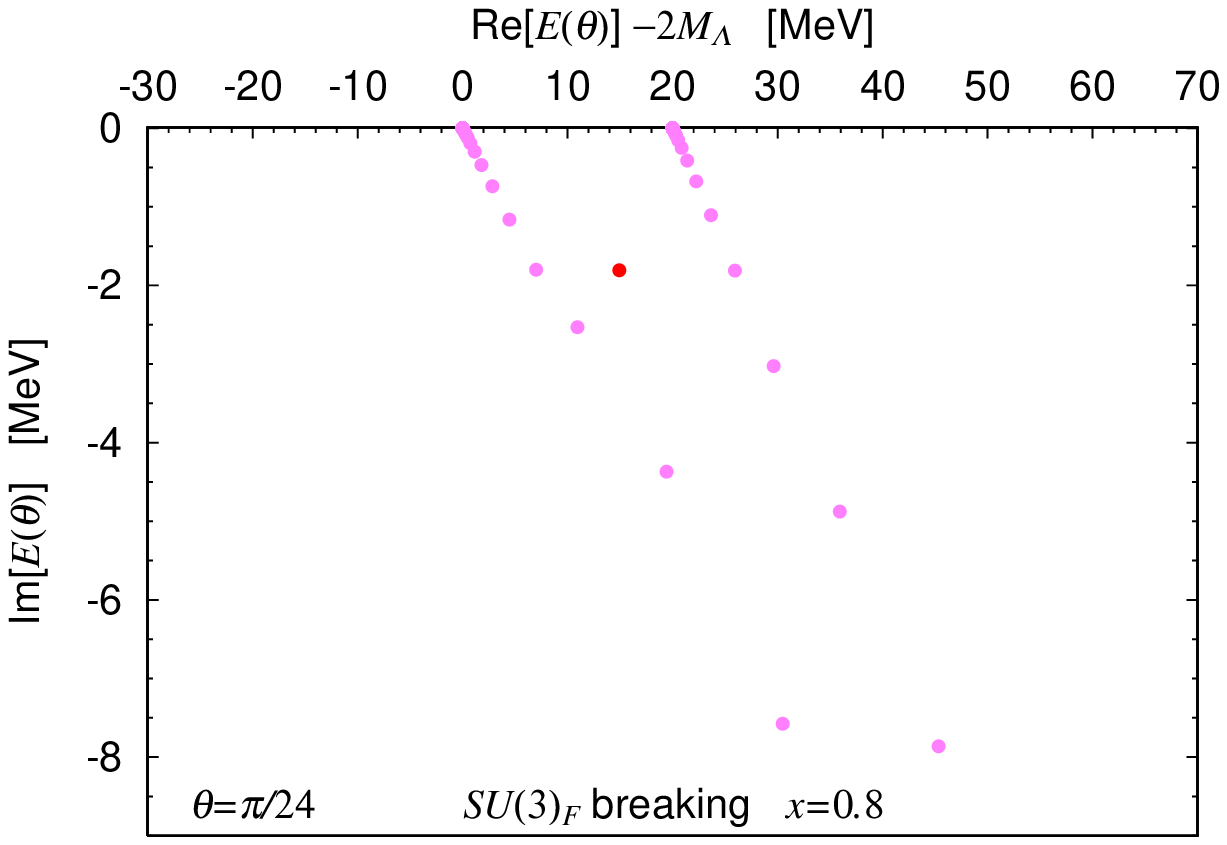} \hfill
\includegraphics[width=0.49\textwidth]{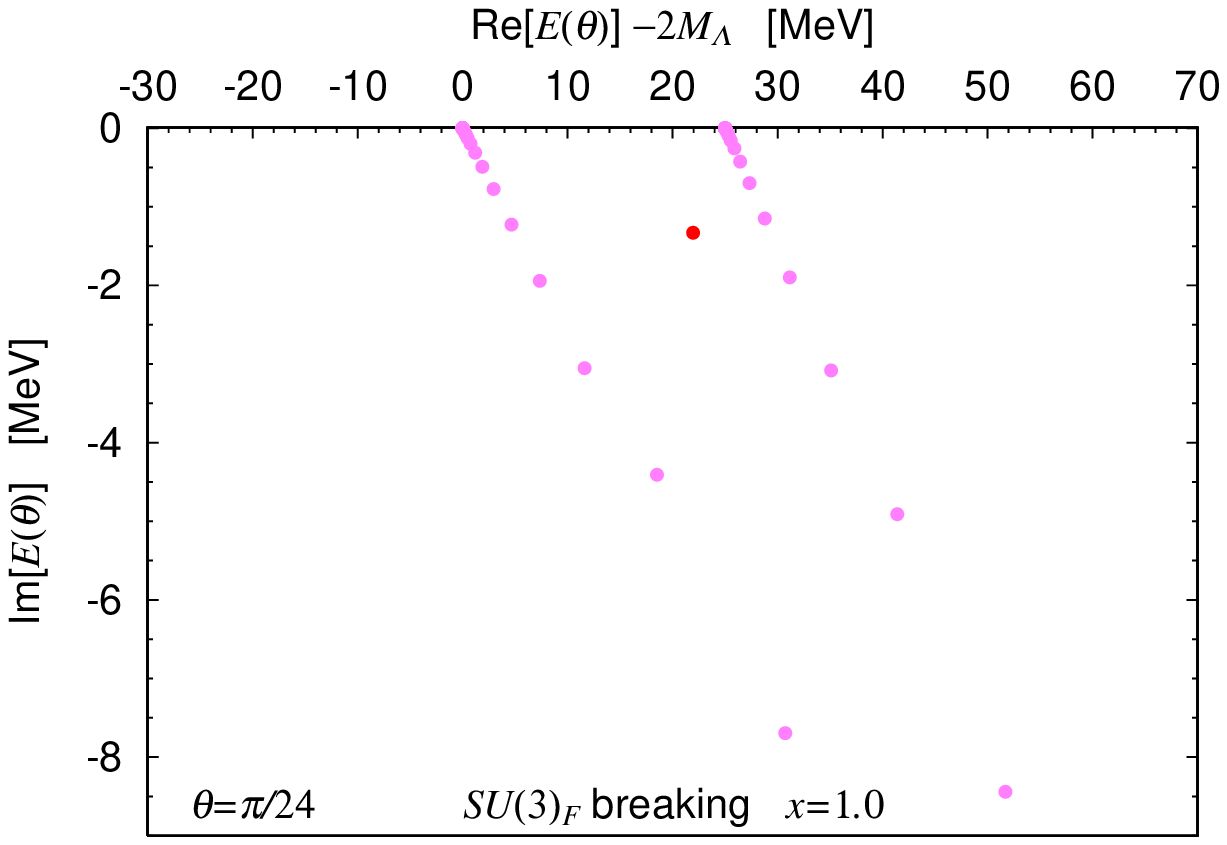}
\smallskip
\caption{Energy eigenvalues of the $\Lambda\Lambda$-$N\Xi$-$\Sigma\Sigma$ coupled system in $^1S_0$ state,
         obtained in the complex scaling method, at several values of the $SU(3)$ breaking parameter $x$.}
\label{fig:CSM}
\end{figure}

The behavior of the $H$ dibaryon (either bound state or resonance) can be seen more directly
by its energy eigenvalue  in the complex scaling method as shown in Fig.~\ref{fig:CSM}.
In this method, Hamiltonian of the system ${\cal H}$ is rotated to ${\cal H}(\theta)$ by ``scaling'' the coordinate $r$ to $r e^{i \theta}$,
so that the eigenvalue equation ${\cal H}(\theta)\Psi_{\theta} = E(\theta) \Psi_{\theta}$
can be easily solved by using a set of square-integrable functions such as the Gaussian function.
There is a theorem, known as the ABC theorem, that an eigenvalue $E(\theta)$ of a bound or resonance state
is independent of the scaling angle $\theta$~\cite{Aguilar:1971ve}.
Aligned dots in Fig.~\ref{fig:CSM} correspond to the continuum of scattering states,
which are discrete  due to a finite number of bases.
We observe one isolated energy eigenvalue on the real axis at $x=0.0$, $0.2$ and $0.4$, corresponding to a bound $H$-dibaryon,
and a complex one between two continua at $x=0.8$ and $1.0$, corresponding to a resonant $H$-dibaryon.
With the physical value of the flavor $SU(3)$ breaking at $x=1$, the $H$-dibaryon exists at 3 MeV below the $N\Xi$ threshold
and has a width of 2.7 MeV in the present estimate. 
Similar results were obtained in phenomenological models~\cite{Oka:1983ku,Mulders:1982da}.

\begin{figure}[p]
\includegraphics[width=0.49\textwidth]{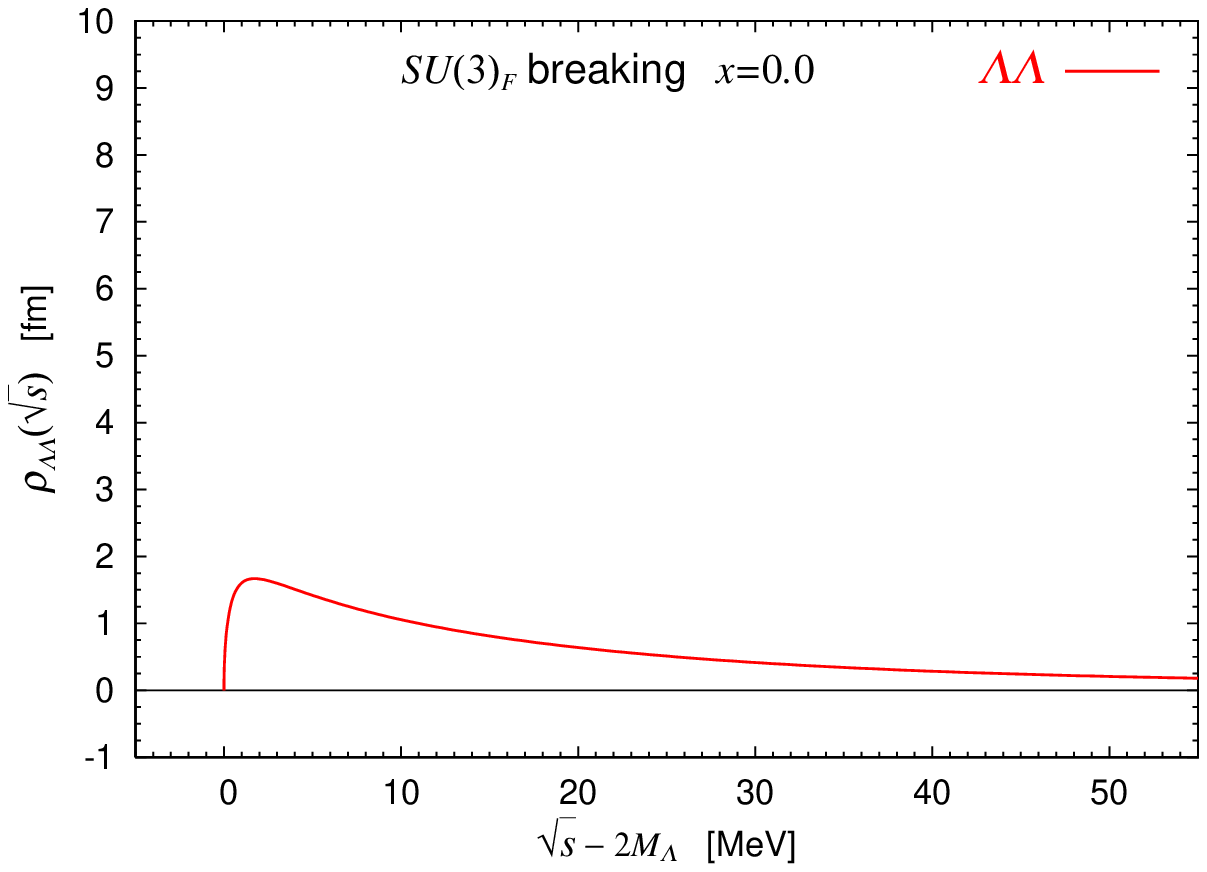} \hfill
\includegraphics[width=0.49\textwidth]{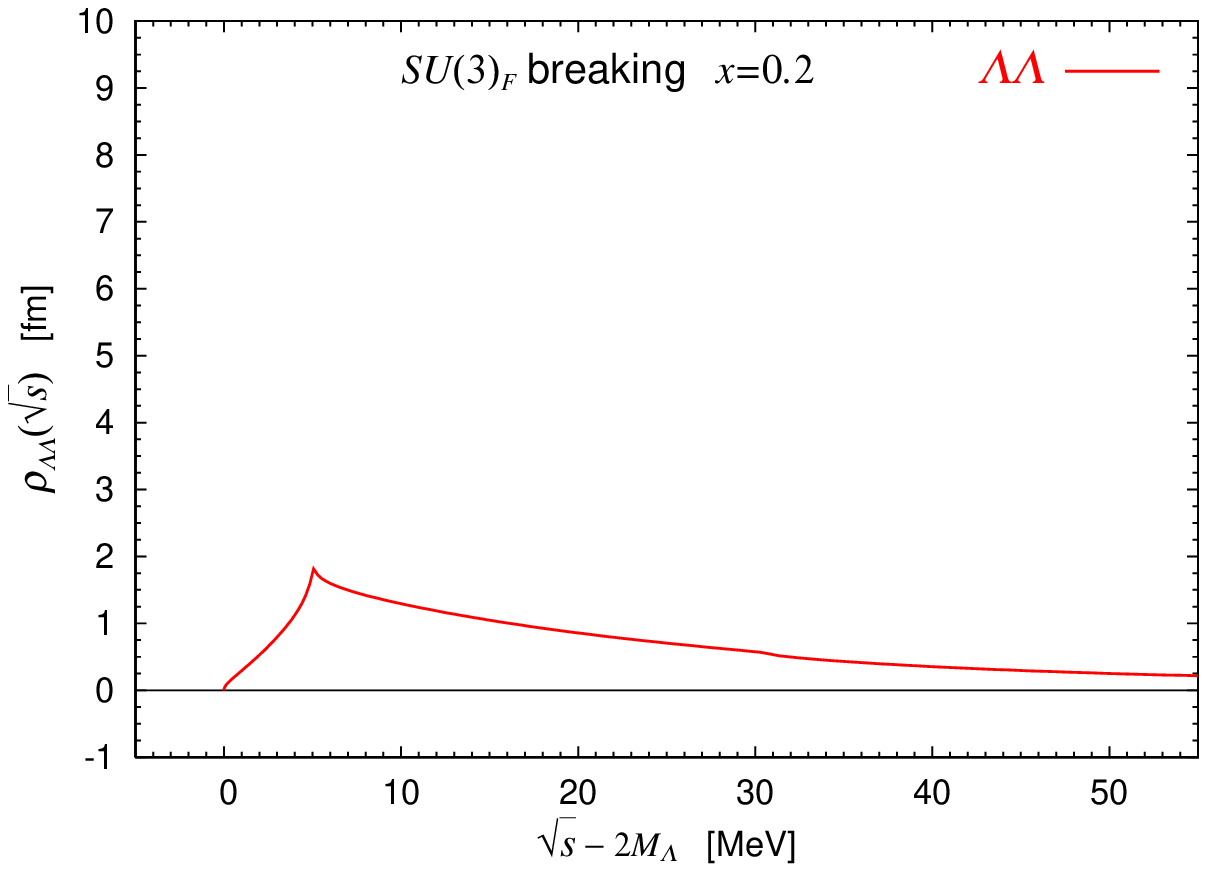}
\smallskip\newline
\includegraphics[width=0.49\textwidth]{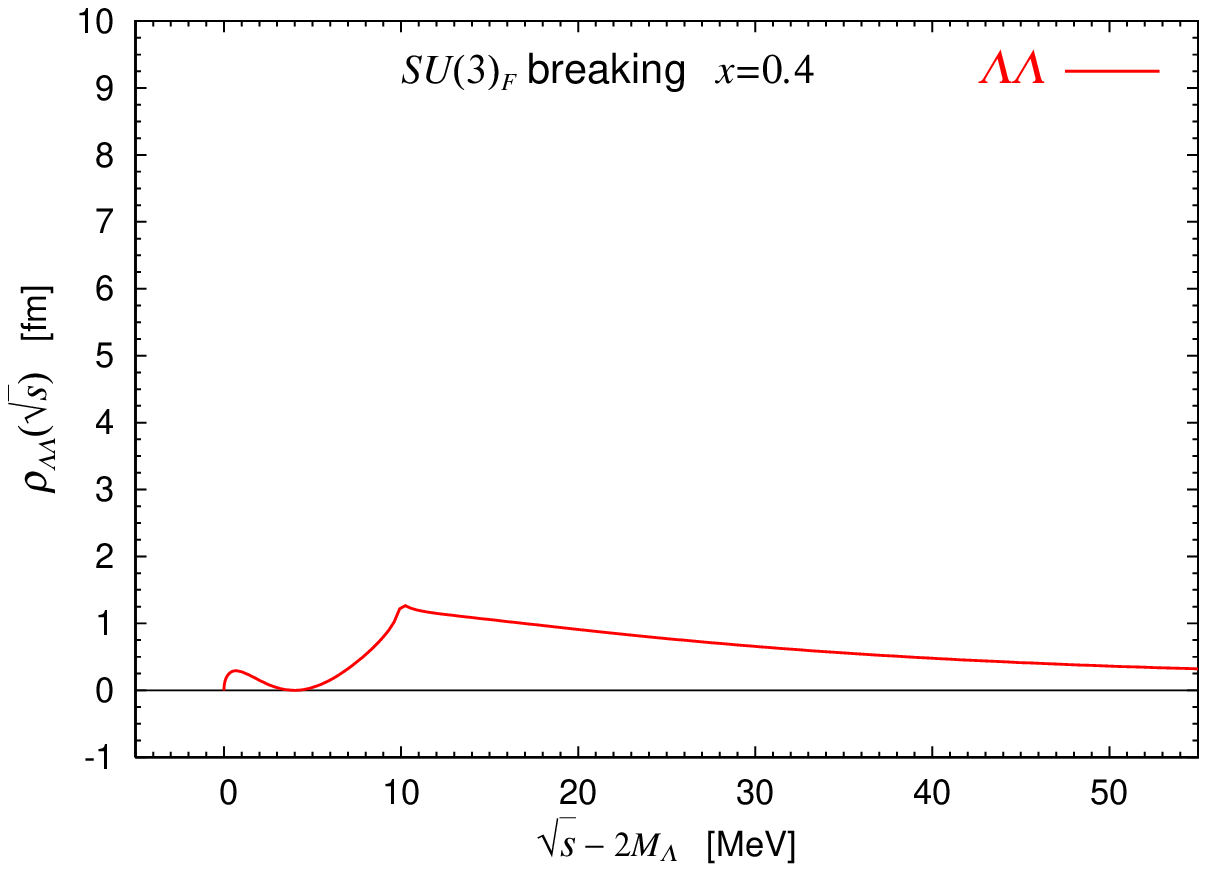} \hfill
\includegraphics[width=0.49\textwidth]{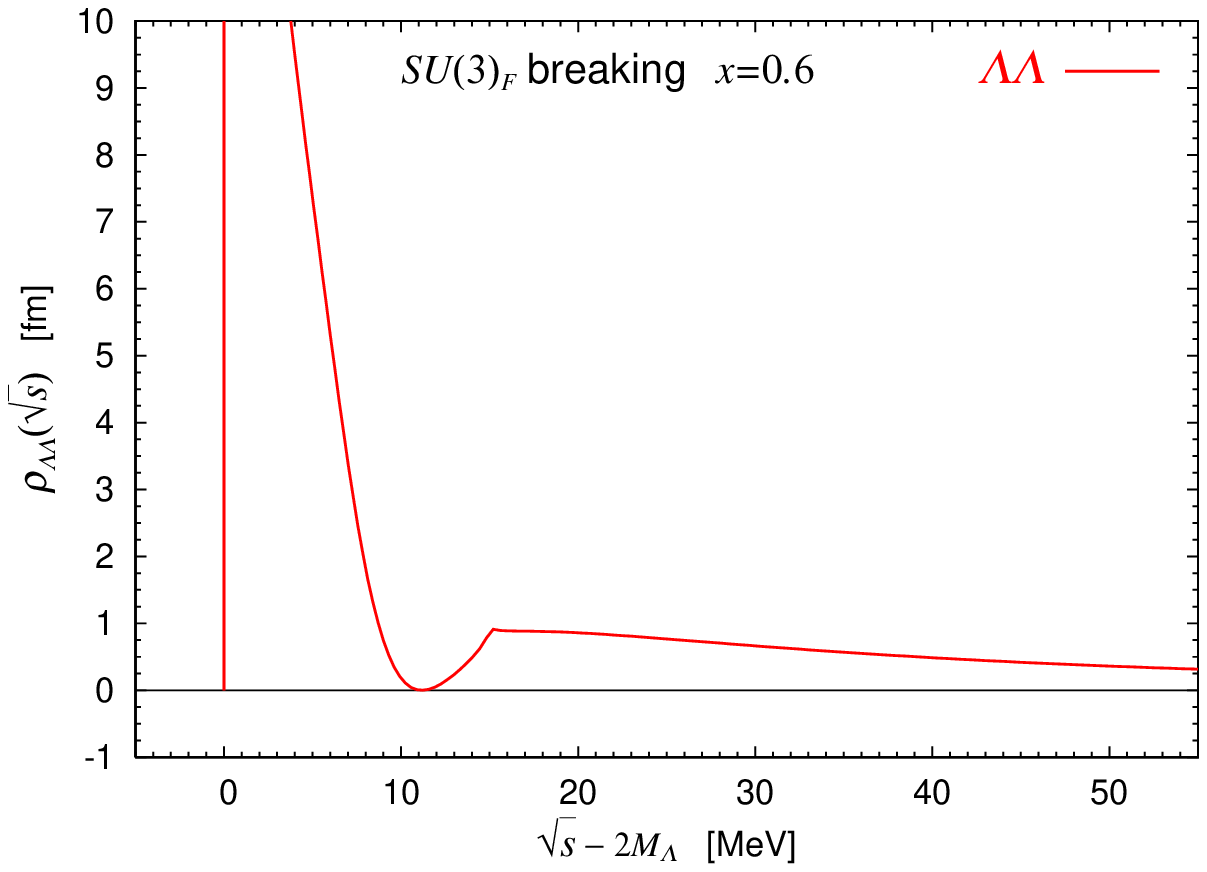}
\smallskip\newline
\includegraphics[width=0.49\textwidth]{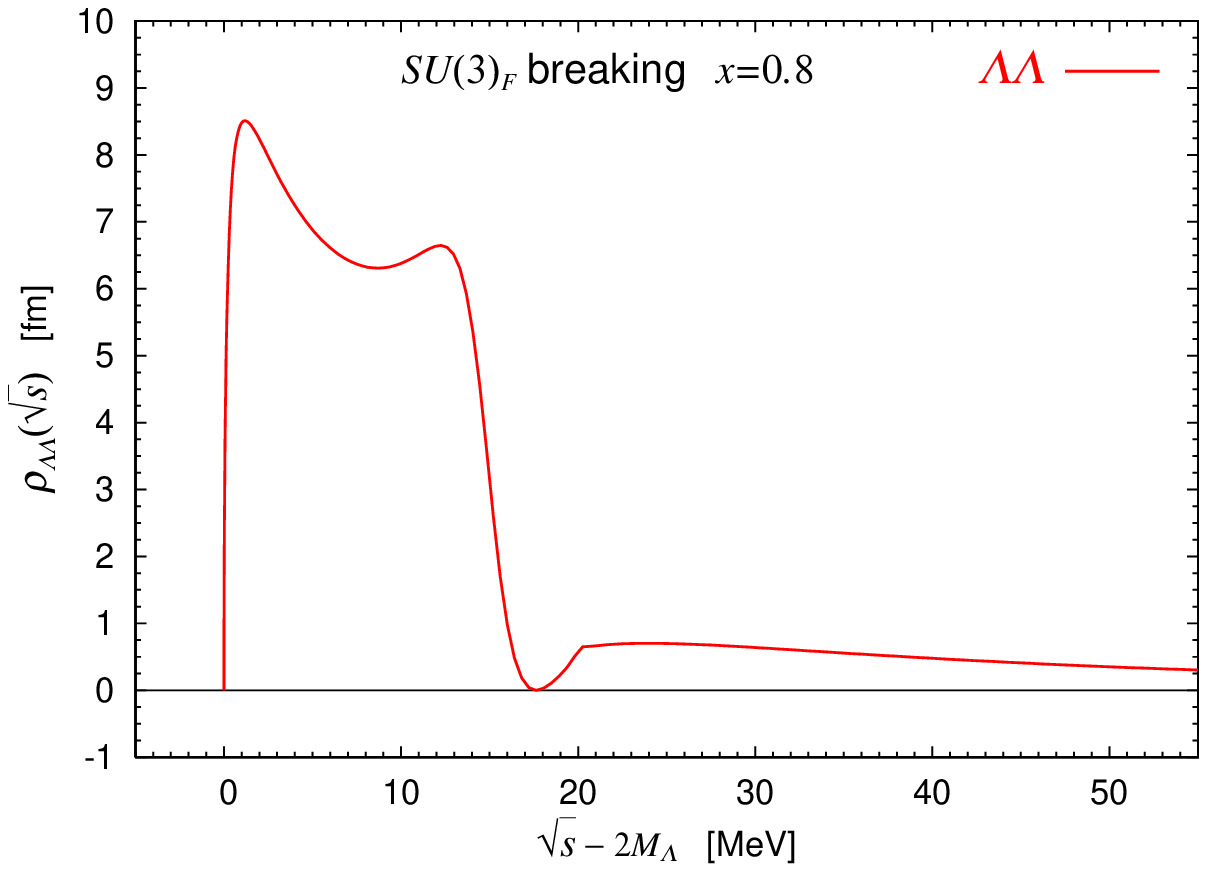} \hfill
\includegraphics[width=0.49\textwidth]{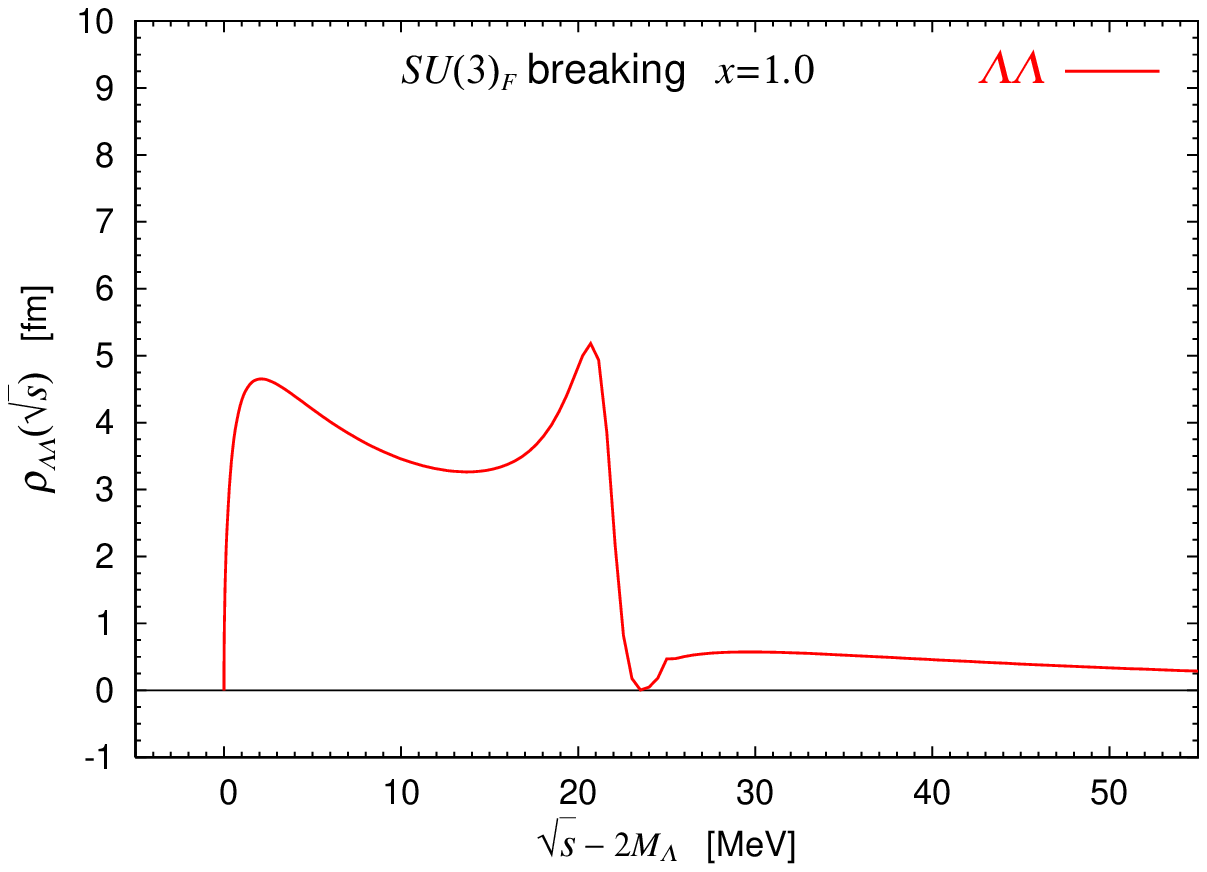}
\smallskip
\caption{The invariant-mass-spectrum of $\Lambda\Lambda$ calculated by assuming the S-wave dominance,
at several values of the $SU(3)$ breaking parameter $x$.
   }
\label{fig:spectrum}
\end{figure}

Shown in Fig.~\ref{fig:spectrum} is the invariant-mass-spectrum of the process $\Lambda\Lambda \to \Lambda\Lambda$ given by 
$\rho_{\Lambda\Lambda}(\sqrt s) = \left|S_{\Lambda\Lambda,\Lambda\Lambda}^{l=0} - 1 \right|^2/k$ with an assumption of S-wave dominance. 
A peak which corresponds to the $H$-dibaryon can be clearly seen at $x=0.6$, $0.8$ and $1.0$. 
This demonstrates that there is a chance for experiments of counting two $\Lambda$'s to confirm the existence of the resonant $H$-dibaryon in nature.
Deeply bound $H$-dibaryon with the binding energy $B_H > 7 $ MeV from the $\Lambda\Lambda$ threshold 
has been ruled out by the discovery of the double $\Lambda$ hypernucleus,
$_{\Lambda \Lambda}^{\ \ 6}$He~\cite{Takahashi:2001nm}. 
On the other hand, an enhancement of the two $\Lambda$'s production has been observed at a little
above the $\Lambda\Lambda$ thresholds in E224 and E522 experiments at KEK,
though statistics is not enough for a definite conclusion~\cite{Yoon:2007aq}.
New data with high statistics from such an experiment at J-PARC~\cite{AhnImai}
as well as the data from heavy-ion experiments at RHIC and LHC~\cite{Shah:2011en} will shed more lights on the nature of the $H$-dibaryon.

\section{Summary}

In this paper, we have
derived baryon-baryon potentials from lattice QCD simulations with flavor $SU(3)$ symmetry.
The lattice volume is taken to be $L \simeq 4$ fm and the pseudo-scalar-meson mass $M_{\rm ps}$ ranges from 469 MeV to 1171 MeV.
Observables such as the  $N\!N$ scattering phase-shifts and the mass of $H$-dibaryon are extracted from  obtained potentials.
The $NN$ phase-shifts (or more precisely the phase-shifts in the flavor 27-plet) shows qualitatively similar behavior with the   
experimental data and the similarity becomes better as the quark mass decreases. 
By solving the few-body Schr\"{o}dinger equation, we found that the present $NN$ potential does not
yet provide bound deuteron nor $^3$He, but there is a shallow bound state for $^4$He  for $M_{\rm ps}=469$ MeV.
  
We have also shown that a stable $H$-dibaryon exists in the flavor-singlet $J^P=0^+$ channel
with the binding energy of about 26 MeV for the lightest quark mass  $M_{\rm ps}=469$ MeV. 
To estimate the effect of flavor $SU(3)$ symmetry breaking on the $H$-dibaryon,
we solve the coupled-channel Schr\"{o}dinger equation  in the $S=-2$ sector ($\Lambda\Lambda$-$N\Xi$-$\Sigma\Sigma$)
by using the baryon masses with approximate $SU(3)$ breaking and the  $BB$ potential in the $SU(3)$ limit.   
We found that the pole position of the $H$-dibaryon crosses the $\Lambda\Lambda$ threshold 
as the flavor $SU(3)$ breaking become larger, and a resonant $H$-dibaryon appears below $N\Xi$ threshold at the physical point.
This is however not a final conclusion due to various approximations about the $SU(3)$ breaking in the Schr\"{o}dinger equation.
Coupled channel (2+1)-flavor simulations with flavor $SU(3)$ breaking are currently underway.
Such lattice QCD simulations together with the laboratory experiments will eventually clarify the nature of the elusive $H$-dibaryon.   

\appendix

\section{Few-nucleon systems from lattice QCD in the flavor SU(3) limit} 
\label{sec:fewnucl}

\begin{figure}[t]
\centering
\includegraphics[width=0.49\textwidth]{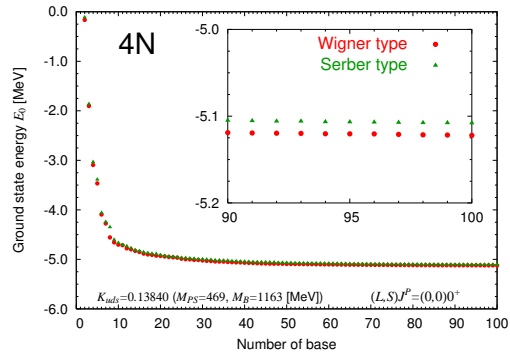}
\caption{Energy of $4N$ ground state 
         in the present flavor $SU(3)$ symmetric world, as a function of a number of bases.
         Two types of two-body interactions are compared. See the text for more details.}
\label{fig:alpha}
\end{figure}

In this appendix, we study few-nucleon systems using the $NN$ potentials obtained in the present calculation.
For simplicity, an effective central potential, which contains a contribution from the tensor potential implicitly,
is employed for the spin-triplet sector, while the central potential is used for the spin-singlet sector.
Since we have determined the leading-order potentials through S and D waves, 
only the parity-even part of potentials is obtained.
We therefore consider two cases: the Wigner type force where the odd part of potentials is equal to the even part, 
and the Serber type force where the odd part is absent.
Since the parity-odd part is known to have little effect to the S-shell nuclei,
our approximation is reasonable at least for three and four nucleon systems.

By searching for the ground state in the stochastic variational method~\cite{Varga:1997xga},
we find a four-nucleon bound state in $(L,S)J^P=(0,0)0^+$ configuration corresponding to the $^4$He nucleus
at the lightest quark mass ($M_{\rm ps}=469$ MeV),
while no three-nucleon bound state is found in the present range of the quark masses.
Fig.~\ref{fig:alpha} shows the energy of the bound state against the number of bases.
Result with the Wigner-type force and that with the Serber-type force agree,
so that the effect of parity-odd potential is small as expected.
We observe that the binding energy of $^4$He is about 5.1 MeV for $M_{\rm ps}=469$ MeV. 
Note that the three-nucleon and four-nucleon forces are absent in the present calculation. 

An indication of very shallow four-nucleon bound state is seen
also for the heaviest quark mass and the second heaviest quark mass
corresponding to $M_{\rm ps}=1171$ MeV and $M_{\rm ps}=1015$ MeV, respectively.
Since the obtained binding energy is very small
(0.8 MeV for $M_{\rm ps}=1171$ MeV and 1.1 MeV for $M_{\rm ps}=1015$ MeV),
and the error of the potentials due to lattice discretization may be large at heavy quark, 
we leave the conclusion on these signals for a future investigation.

\section*{Acknowledgment}
We thank K.-I. Ishikawa and PACS-CS group for providing their DDHMC /PHMC code~\cite{Aoki:2008sm},
and authors and maintainer of CPS++~\cite{CPS}, whose modified version is used in this paper.
Numerical computations of this work have been carried out at Univ.~of Tsukuba supercomputer system (T2K).
This research is supported in part by Grant-in-Aid
for Scientific Research on Innovative Areas(No.2004:20105001, 20105003) and
for Scientific Research(C) 23540321, JSPS 21$\cdot$5985 and SPIRE (Strategic Program for Innovative REsearch).



\bibliographystyle{elsarticle-num}

\begin{thebibliography}{99}

\bibitem{Hashimoto:2006aw} 
  O.~Hashimoto and H.~Tamura,
  Prog.\ Part.\ Nucl.\ Phys.\  {\bf 57}, 564 (2006).
\bibitem{SchaffnerBielich:2010am} 
  J.~Schaffner-Bielich,
  Nucl.\ Phys.\ A {\bf 835}, 279 (2010)
  [arXiv:1002.1658 [nucl-th]].
  
\bibitem{Jaffe:1976yi}
  R.~L.~Jaffe,  
  Phys.\ Rev.\ Lett.\  {\bf 38}, 195 (1977)
  [Erratum-ibid.\  {\bf 38}, 617 (1977)].

\bibitem{Sakai:1999qm}
  T.~Sakai, K. Shimizu and K.~Yazaki,
  Prog.~Theor.~Phys.~Suppl. {\bf 137}, 121 (2000)
  [arXiv:nucl-th/9912063].
  
\bibitem{Takahashi:2001nm}
  H.~Takahashi {\it et al.},
  Phys.\ Rev.\ Lett.\  {\bf 87}, 212502 (2001).

\bibitem{Yoon:2007aq}
  C.~J.~Yoon {\it et al.},
  Phys.\ Rev.\  C {\bf 75}, 022201 (2007).  
  

\bibitem{Luscher:1990ux}
  M.~L\"{u}scher,
  Nucl.\ Phys.\  B {\bf 354}, 531 (1991).

\bibitem{Ishii:2006ec}
  N.~Ishii, S.~Aoki and T.~Hatsuda,
  Phys.\ Rev.\ Lett.\  {\bf 99}, 022001 (2007)  [arXiv:nucl-th/0611096].
\bibitem{Aoki:2009ji}
  S.~Aoki, T.~Hatsuda and N.~Ishii,
  Prog.\ Theor.\ Phys.\  {\bf 123}, 89 (2010)  [arXiv:0909.5585 [hep-lat]].


\bibitem{Fukugita:1994ve}
  M.~Fukugita {\it et al.},
  Phys.\ Rev.\  D {\bf 52}, 3003 (1995) [arXiv:hep-lat/9501024].
\bibitem{Beane:2002nu}
  S.~R.~Beane and M.~J.~Savage,
  Phys.\ Lett.\  B {\bf 535}, 177 (2002)  [arXiv:hep-lat/0202013].
\bibitem{Beane:2010em}
  S.~R.~Beane, W.~Detmold, K.~Orginos and M.~J.~Savage,
  Prog.\ Part.\ Nucl.\ Phys.\  {\bf 66}, 1 (2011)
  [arXiv:1004.2935 [hep-lat]].


\bibitem{Nemura:2008sp}
  H.~Nemura, N.~Ishii, S.~Aoki and T.~Hatsuda,
  Phys.\ Lett.\  B {\bf 673}, 136 (2009)
  [arXiv:0806.1094 [nucl-th]].
\bibitem{Inoue:2010hs}
  T.~Inoue {\it et al.}  [HAL QCD Coll.],
  Prog. Theor. Phys. {\bf 124}, 591 (2010)
  [arXiv:1007.3559 [hep-lat]].
  \bibitem{Inoue:2010es} 
  T.~Inoue {\it et al.}  [HAL QCD Collaboration],
  Phys.\ Rev.\ Lett.\  {\bf 106}, 162002 (2011)
  [arXiv:1012.5928 [hep-lat]].
\bibitem{Ikeda:2011qm} 
  Y.~Ikeda [for HAL QCD Collaboration],
  arXiv:1111.2663 [hep-lat].
\bibitem{Kawanai:2010ru} 
  T.~Kawanai and S.~Sasaki,
  PoS LATTICE {\bf 2010}, 156 (2010)
  [arXiv:1011.1322 [hep-lat]].

  

 \bibitem{Beane:2010hg}
  S.~R.~Beane {\it et al.}  [NPLQCD Collaboration],
  Phys.\ Rev.\ Lett.\  {\bf 106}, 162001 (2011)
  [arXiv:1012.3812 [hep-lat]].


\bibitem{Ishii:2011}
  N.~Ishii {\it et al.}  [HAL QCD Coll.], in preparation

\bibitem{Inoue:2011tk}
  T.~Inoue  [for HAL QCD Collaboration],
  arXiv:1111.5098 [hep-lat].


  \bibitem{Beane:2011xf} 
  S.~R.~Beane, E.~Chang, W.~Detmold, B.~Joo, H.~W.~Lin, T.~C.~Luu, K.~Orginos and A.~Parreno,
  Mod.\ Phys.\ Lett.\ A {\bf 26}, 2587 (2011)
  [arXiv:1103.2821 [hep-lat]].
 \bibitem{Shanahan:2011su} 
  P.~E.~Shanahan, A.~W.~Thomas and R.~D.~Young,
  Phys.\ Rev.\ Lett.\  {\bf 107}, 092004 (2011)
  [arXiv:1106.2851 [nucl-th]];
  A.~W.~Thomas, P.~E.~Shanahan and R.~D.~Young,
  arXiv:1111.0114 [nucl-th].     
 \bibitem{Haidenbauer:2011ah} 
  J.~Haidenbauer and U.~-G.~Meissner,
  Phys.\ Lett.\ B {\bf 706}, 100 (2011)
  [arXiv:1109.3590 [hep-ph]]; and in Nucl. Phys. A, this Volume (2012)  [arXiv:1111.4069 [nucl-th]] 

\bibitem{Oka:1983ku}
  M.~Oka, K.~Shimizu and K.~Yazaki,
  Phys.\ Lett.\  B {\bf 130}, 365 (1983).

\bibitem{Aoki:2011gt} 
  S.~Aoki {\it et al.}  [HAL QCD Collaboration],
 Proc. Jpn. Acad., Ser. B, {\bf 87} 509-517 (2011)  [arXiv:1106.2281 [hep-lat]].  
  
\bibitem{Murano:2011nz}
  K.~Murano, N.~Ishii, S.~Aoki and T.~Hatsuda,
  Prog.\ Theor.\ Phys.\  {\bf 125}, 1225 (2011)
  [arXiv:1103.0619 [hep-lat]].

\bibitem{Iwasaki}
Y.~Iwasaki, 
arXiv:1111.7054[hep-lat].

\bibitem{CPPACS-JLQCD}
CP-PACS and JLQCD Coll., http://\\
www.jldg.org/ildg-data/CPPACS+JLQCDconfig.html

\bibitem{sasaki2010} 
  K.~Sasaki [HAL QCD Coll.], PoS {\bf LAT 2010}, 157 (2010).

\bibitem{NN-OnLine}
  NN-OnLine, http://nn-online.org

\bibitem{Yamazaki:2011nd}
  T.~Yamazaki, Y.~Kuramashi, A.~Ukawa for the PACS-CS Collaboration,
  arXiv:1105.1418 [hep-lat].
\bibitem{Beane:2011iw}
  S.~R.~Beane {\it et al.}  [NPLQCD Collaboration],
  arXiv:1109.2889 [hep-lat].
  

\bibitem{Aguilar:1971ve}
  J.~Aguilar and J.~M.~Combes,
  Commun.\ Math.\ Phys.\  {\bf 22}, 269 (1971).
  E.~Balslev and J.~M.~Combes,
  Commun.\ Math.\ Phys.\  {\bf 22}, 280 (1971).

\bibitem{Mulders:1982da}
  P.~J.~Mulders and A.~W.~Thomas,
  J.\ Phys.\ G {\bf 9}, 1159 (1983).

\bibitem{AhnImai}
 J.K.Ahn and K.Imai; J-PARC proposal P42 

\bibitem{Shah:2011en} 
  N.~Shah [for the STAR Collaboration],
  arXiv:1112.0590 [hep-ex].
   
\bibitem{Varga:1997xga}
  K.~Varga and Y.~Suzuki,
  Comput.\ Phys.\ Commun.\  {\bf 106}, 157 (1997)
  [arXiv:nucl-th/9702034].

\bibitem{Aoki:2008sm}
  S.~Aoki {\it et al.}  [PACS-CS Coll.],
  Phys.\ Rev.\  D {\bf 79}, 034503 (2009)
  [arXiv:0807.1661 [hep-lat]].
  
\bibitem{CPS}
  Columbia Physics System (CPS), http://qcdoc.phys.columbia.edu/cps.html
  
\end{thebibliography}







\end{document}